\documentclass[journal]{IEEEtran}
\usepackage{graphicx}
\usepackage{graphics}
\usepackage{mathptmx} 
\usepackage{times} 
\usepackage{mathdots}
\usepackage{array}
\usepackage{amsmath, amsthm, amssymb, fancyhdr, lastpage}
\usepackage{nicefrac}
\usepackage{epsfig} 
\usepackage{epstopdf} 
\usepackage{cite}        
\usepackage{tabularx,ragged2e,booktabs}
\usepackage{enumitem}
\newcommand*\mean[1]{\overline{#1}}

\begin{document}

\title{A Dynamical Systems Approach to Modeling and Analysis of Transactive Energy Coordination}

\author{Md~Salman~Nazir,~\IEEEmembership{Member,~IEEE,}
        and~Ian~A.~Hiskens,~\IEEEmembership{Fellow,~IEEE}
\thanks{Authors are with the Department of Electrical Engineering and
	Computer Science, University of Michigan, Ann Arbor, MI, USA. \, Emails: mdsnazir@umich.edu, hiskens@umich.edu.}

\thanks{The authors gratefully acknowledge the
	contribution of the Natural Sciences and Engineering Research
	Council of Canada (NSERC) and the U.S.~National Science Foundation
	through grant CNS-1238962.}
}

\maketitle

\begin{abstract}
Under transactive (market-based) coordination, a population of
distributed energy resources (DERs), such as thermostatically
controlled loads (TCLs) and storage devices, bid into an energy
market. Consequently, a certain level of demand will be cleared based
on the operating conditions of the grid. This paper analyzes the
influence of various factors, such as price signals, feeder limits,
and user-defined bid functions and preferences, on the aggregate
energy usage of DERs. We identify cases that can lead to load
synchronization, undesirable power oscillations and highly volatile
prices. To address these issues, the paper develops an aggregate model
of DERs under transactive coordination. A set of Markov transition
equations have been developed over discrete ranges (referred to as
``bins'') of price levels and their associated DER operating states. A
detailed investigation of the performance of this aggregate model is
presented. With reformulation of the transition equations, the bin
model has been incorporated into a model predictive control setting
using both mixed integer programming and quadratic programming. A case
study shows that a population of TCLs can be managed economically
while avoiding congestion in a distribution grid. Simulations also
demonstrate that power oscillations arising from synchronization of
TCLs can be effectively avoided.
\end{abstract} 


%
\IEEEpeerreviewmaketitle

\section{Introduction}

Transactive energy coordination mechanisms have been proposed as a
framework for managing large numbers of distributed energy resources
(DERs), such as thermostatically controlled loads (TCLs), energy
storage and electric vehicles, in an electric grid
\cite{Huang2010,Fuller2011,Li2016a}. Such schemes provide users with
the flexibility to consume energy based on the price they are willing
to pay. Several recent studies and demonstration projects have shown
the applicability of such mechanisms to manage the aggregate demand of
residential electric loads and commercial building heating,
ventilation, and air conditioning (HVAC) systems \cite{Huang2010,
	Li2016a, Li2016b, Fuller2011, Knudsen2016, Hao2016}. Applications
include reduction of peak demand, provision of regulation services and
congestion relief.

Under transactive (market-based) energy coordination, a population of
distributed resources bids into the energy market and a certain level
of demand is cleared, depending on the operating conditions of the
grid. This process is primarily based on a double auction mechanism
\cite{Fuller2011}. A market mechanism proposed in \cite{Li2016a} can
attain economically efficient market outcomes while taking into
account individual device dynamics. The applicability of this approach
and the influence of different system parameters have been studied in
\cite{Li2016b}. While these market-based schemes can attain
economically efficient outcomes, concerns remain regarding the impact
on system stability. Previous analyses of electricity market dynamics
\cite{Alvarado1999, Roozbehani2012} have highlighted the possibility
of power oscillations and highly volatile prices. Our recent work
\cite{Nazir2017c} applied the bidding strategy described in
\cite{Fuller2011} (also in Section~II of \cite{Li2016a}) to show
undesirable oscillatory response of a population of
air-conditioners. Simulations suggest that several factors may
contribute to load synchronization and power oscillations, including
prolonged flat price signals followed by sharp changes in the price,
feeder limits that are set too low and that are encountered
periodically, and the form of user bid curves. A detailed analysis of
emergent oscillations was beyond the scope of \cite{Nazir2017c} and
will be elaborated upon in this paper.

It has been shown in \cite{Callaway2011} and \cite{Zhao2017} that
oscillations may appear under market-based coordination of DERs. In
\cite{Callaway2011}, oscillatory behavior was observed in the
aggregate demand when simulating a large number of price-responsive
electric vehicles. It was shown in \cite{Zhao2017} that while
oscillations may be present, the aggregate demand from a population of
DERs can reach the stable equilibrium provided the user-defined bid
slopes are not too steep. This work considered a generalized aggregate
battery model. Our proposed framework also builds upon a generalized
battery model. Additionally, we introduce the concept of
\textit{lockout conditions} to account for customers who desire
periods of uninterrupted supply. The possibility and nature of
oscillations in the aggregate demand, given the presence of such
lockout conditions, will be studied.


Prior analysis has clearly demonstrated that the aggregate demand of
DERs can be affected by price signals, feeder capacity limits, and DER
bidding strategies. However, predicting the aggregate behavior of DERs
under transactive energy coordination and guaranteeing stability
remains a challenging task \cite{Nazir2017c, Zhao2017}. Therefore,
this paper focuses largely on developing an aggregate model of DERs
under market-based coordination. To accomplish this goal, a set of
Markov transition equations are developed over discrete ranges of DER
energy and price levels (here referred to as ``bins''). Lockout
conditions are also incorporated. Eigenvalues of the resulting system
matrix provide insights into system behavior. Although bin-based
aggregate models have been used extensively for controlling
aggregations of TCLs \cite{Koch2011, Mathieu2013, Totu2016,
  Nazir2016}, application to the transactive coordination framework
has not previously been considered. Hence, a novel contribution of
this paper is the extension of the bin-modeling approach to
market-based coordination schemes for DERs. Furthermore, incorporating
this model in a model predictive control (MPC) framework allows the
calculation of optimal price signals for achieving desired DER energy
usage. A reformulation of the transition equations ensures that the
modified model remains linear for optimization-based market clearing
strategies. However, a set of logical constraints are needed to model
the market clearing behavior. Thus, a mixed integer formulation is
obtained, providing a further contribution to aggregate modeling and
market-based coordination of DERs. Additionally, relaxation of the
integer model results in a quadratic program that is computationally
more efficient. Measures to avoid synchronization are also
incorporated in the MPC strategy.

Finally, we consider an application of MPC to manage a population of
DERs in a distribution system. Several recent studies have shown that
with increased penetration of controllable loads and storage devices,
advanced congestion management techniques will be necessary to avoid
simultaneous consumption from DERs and to limit payback from
unsatisfied demand \cite{Verzijlbergh2014, Dubey2015, Hanif2017a,
  Hanif2017b, Ilic2011a}. Hence, we show that a population of TCLs can
be managed economically while avoiding congestion in a distribution
grid. Simulations also demonstrate that synchronization of TCLs and
undesired power oscillations can be effectively avoided. The MPC
formulation considers system level costs, operational constraints and
constraints related to the DER population. Consequently, the generated
price signals induce desirable behavior of the DER population. This
addresses an important gap in transactive coordination as previous
studies \cite{Huang2010,Fuller2011,Li2016a, Li2016b} have taken price
signals to be exogenous inputs rather than generated via feedback.

The remainder of the paper is organized as
follows. Section~\ref{sec:Model} describes the modeling background,
and presents analysis of the behavior of DERs under transactive
coordination. Section~\ref{sec:agg} presents the bin-based aggregate
model for DERs under market-based coordination. Section~\ref{sec:MPC}
presents the model predictive control (MPC) framework for controlling
the DERs. Both a mixed-integer formulation and a relaxed quadratic
program formulation are considered. Section~\ref{sec:Results} provides
simulation results showing the performance of the aggregate model and
the MPC controller. Section~\ref{sec:concl} concludes by summarizing
our findings and discussing their implications.


\section{Problem Formulation and Analysis}\label{sec:Model}

\subsection{Single DER Dynamic Model}

The state dynamics of DERs, such as TCLs, electric vehicles and
batteries, can be expressed using a generalized battery model
\cite{mathieu15a,Hao2013}. Let $e_{j,k}$ be the $j$-th DER's energy
state, i.e. its state of charge (SOC), at time $k$. Then, the
evolution in $e_{j,k}$ can be modeled with the discrete-time
difference equation,
\begin{equation}\label{eq:GBM0}
e_{j, k+1} = a_j e_{j,k} + {d}_{j,k} , 
\end{equation}
where $e_j^{\textrm{min}} \le e_{j,k} \le e_j^{\textrm{max}}$, and
${d}_{j,k}$ is the power consumed, with ${d}_j^{\textrm{min}} \le
{d}_{j,k} \le {d}^{\textrm{max}}$. The coefficient $a_j$ is the $j$-th
DER's energy dissipation rate, with $a_j \in (0,1]$ for TCLs and HVAC
systems, and $a_j \approx 1$ for storage devices \cite{Zhao2017}.
Assume $e_{j,k}$ is normalized, i.e. $e_j^{\textrm{min}} = 0$ and
$e_j^{\textrm{max}} = 1$. Considering $u_{j,k} \in \{0, 1\}$ is the
discrete power on/off decision, and $\gamma_j$ is the charging
efficiency (i.e. the energy gain if the DER is on), we can write,
\begin{equation}\label{eq:GBM1}
e_{j, k+1} = a_j e_{i,k} + \gamma_j u_{j,k}. 
\end{equation}

The specifics on temperature dynamics of TCLs, along with example
parameter values, are provided in Appendix~\ref{sec:TCLmodel}.

Next, let $\pi_{j,k}$ be the $j$-th DER's bid price at time
$k$. Typically $\pi_{j,k}$ decreases as $e_{j,k}$
increases. Therefore, the bid function relating $\pi_{j,k}$ and
$e_{j,k}$ can be expressed as,
\begin{equation}\label{eq:SOCvsBid}
\pi_{j,k} = \pi^{\textrm{max}}_{j} - \beta_j e_{j,k}, 
\end{equation}
where $\pi^{\textrm{max}}_{j}$ is the bid price when $e_{j,k}$ = 0 and
$\beta_j$ represents the slope of the bid curve.

\subsection{Market Clearing Mechanism}\label{sec:MarketBasic}

\begin{figure}[t]
	\begin{center}
		\includegraphics[scale=0.37]{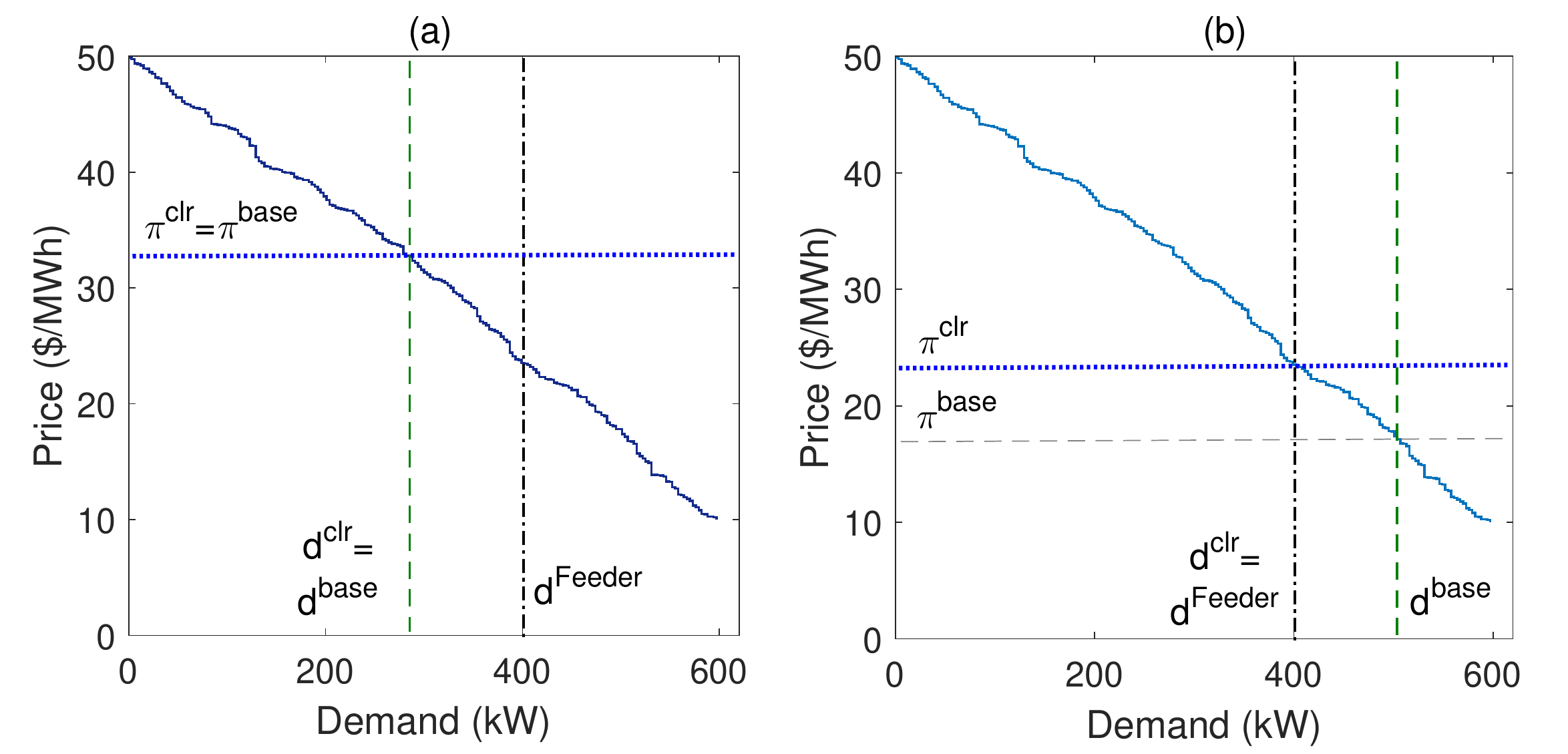}
		\caption{(a) Market clearing with feeder capacity not exceeded. (b)
			Market clearing with feeder capacity exceeded \cite{Fuller2011}, \cite{Nazir2017c}.} \label{fig:TCclearing1}
	\end{center}
\end{figure}

Several different market-based coordination mechanisms have been
proposed in the literature \cite{Fuller2011}, \cite{Li2016a},
\cite{Zhou2017}. For analysis purposes, we primarily focus on the
mechanism proposed in \cite{Fuller2011} (also given in Section~II of
\cite{Li2016a}) and suggest generalizations in later sections.  To
summarize, the DERs are assumed to be connected to a distribution
feeder, which clears an allowable demand level at a particular
price. Let $\pi^{\textrm{base}}_{k}$ be the base price forecast at the
$k$-th time period. An aggregator gathers anonymous bids (price versus
demand) and builds an aggregate demand function, as shown in
Fig.~\ref{fig:TCclearing1}. Using the aggregate demand function and
the base price information $\pi^{\textrm{base}}_k$ for that time
period, the corresponding base aggregate demand $d^{\textrm{base}}_k$
is determined. If $d^{\textrm{base}}_k < d^{\textrm{Feeder}}$ (see
Fig.~\ref{fig:TCclearing1}(a)) then $d^{\textrm{clr}}_{k}$ =
$d^{\textrm{base}}$ and $\pi^{\textrm{clr}}_k$ =
$\pi^{\textrm{base}}_{k}$. If $d^{\textrm{base}}_k \ge=
d^{\textrm{Feeder}}$ (see Fig.~\ref{fig:TCclearing1}(b)) then
$d^{\textrm{clr}}_{k}$ = $d^{\textrm{Feeder}}$ and set
$\pi^{\textrm{clr}}_k$ accordingly.
%
%
%
%
%
%


The response (dispatch decision) of an individual DER to a transactive
incentive signal (price signal) $\pi^{\textrm{clr}}_k$ is given by,
\begin{equation}\label{eq:vclear}
v_{j,k} =
\begin{cases}
0, & \text{if} \quad  \pi_{j,k}  < \pi^{\textrm{clr}}_k, \\
1, & \text{if} \quad  \pi_{j,k} \ge \pi^{\textrm{clr}}_k. \\
\end{cases}
\end{equation}
In this case of market-driven decisions, $u_{j,k}$ can be replaced by
$v_{i,k}$ in \eqref{eq:GBM1}.

\subsection{DER Switching Logic including Lockout Constraint}

\begin{figure}[t!]
	\begin{center}
		\includegraphics[scale=0.335]{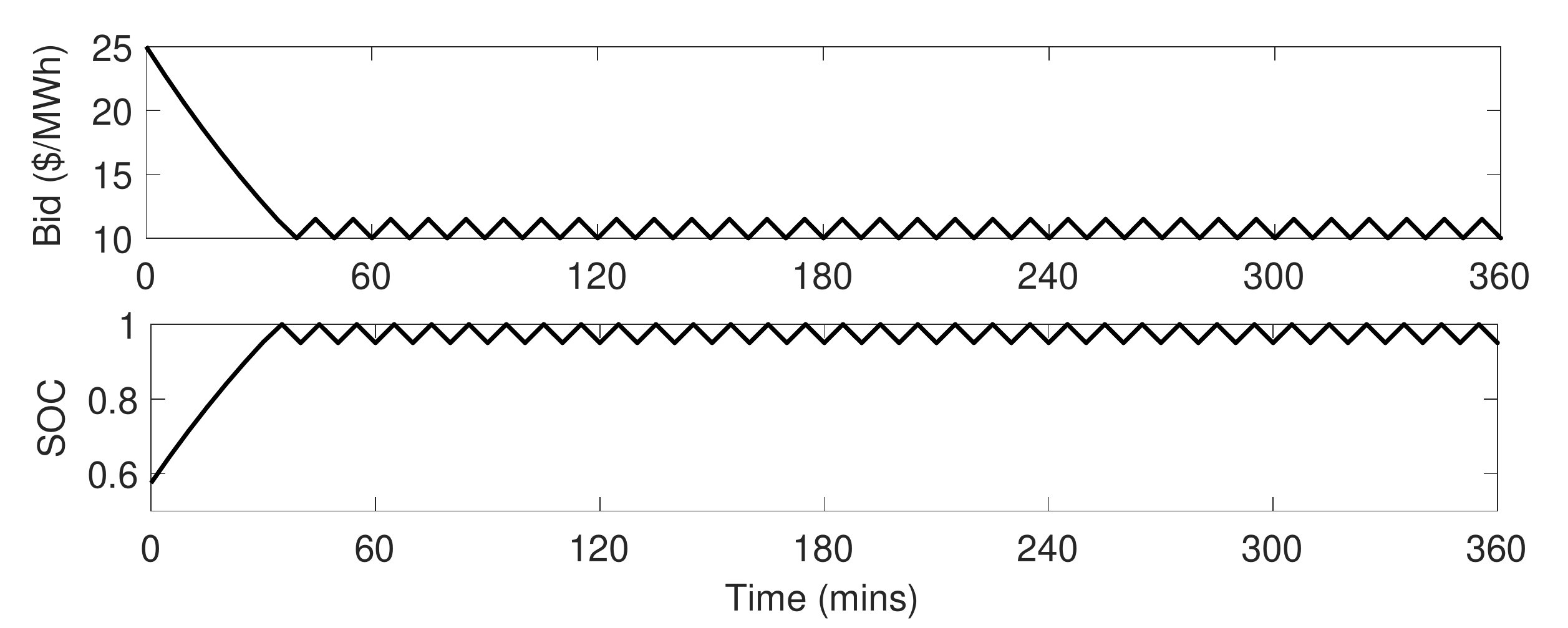}
		\caption{Fast cycling when no lockout mode is present.}
		\label{fig:NoLockout}
	\end{center}	
\end{figure}

In \eqref{eq:GBM1} and \eqref{eq:vclear}, no restrictions have been
imposed on how frequently $u_{j,k}$ can switch on/off. This may lead
to fast cycling, which is often undesirable. Fig.~\ref{fig:NoLockout}
provides an illustration where a DER's bid price was initially above
the clearing price so it charged. With time, the SOC increases and its
bid price decreases. Eventually the DER will reach full SOC or its bid
price will drop below the market clearing price. When that occurs, the
DER switches off, which caused its SOC to drop and bid price to
increase. Thus, its SOC and bid price oscillate within a narrow
region.
\begin{figure}[t!]
	\begin{center}
		\includegraphics[scale=0.335]{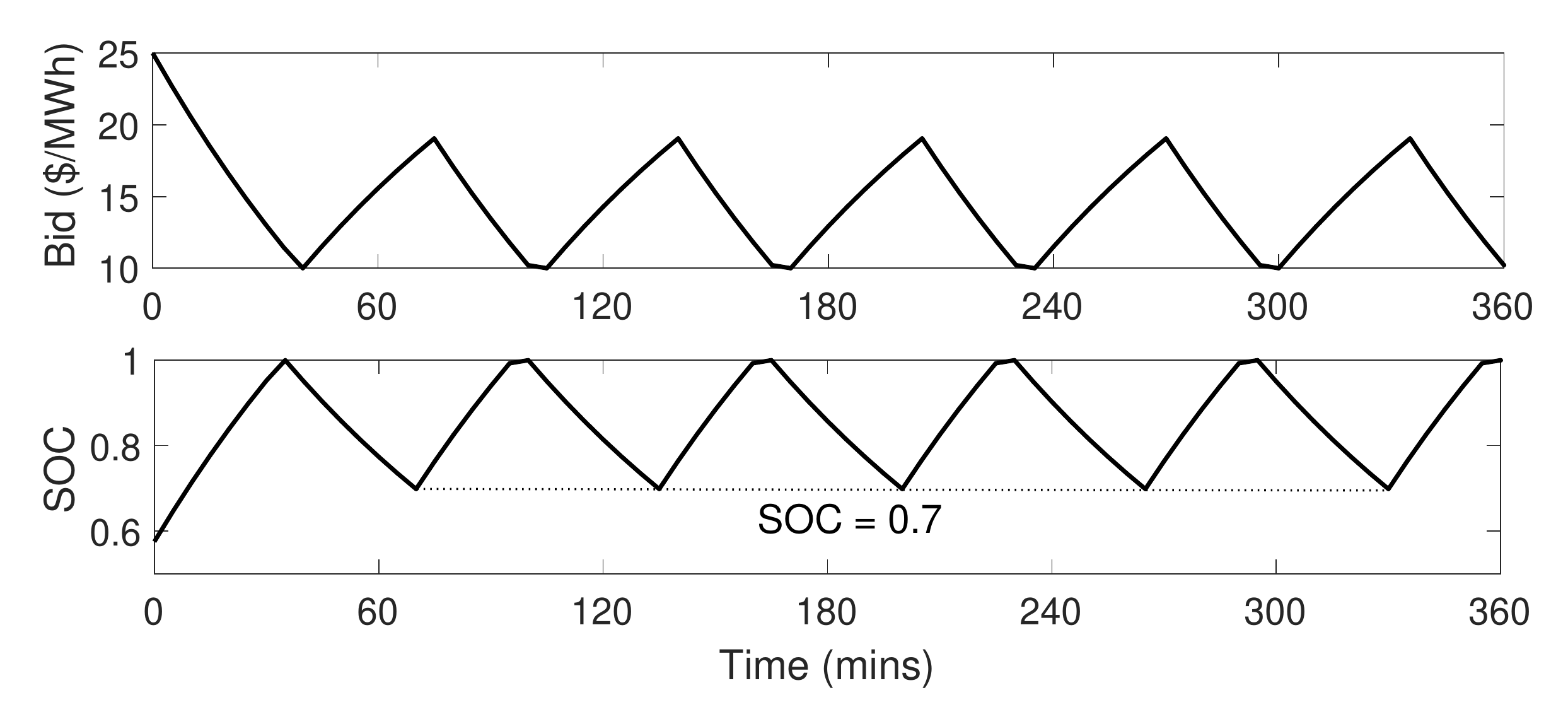}
		\caption{Prolonged on/off periods in the presence of a lockout.}	
		\label{fig:lockout}	
	\end{center}	
\end{figure}
To avoid fast cycling when a DER is already fully charged, lockout
constraints can be incorporated in the DER dispatch logic. For
example, residential air conditioners employ hysteresis control to
prevent fast cycling \cite{Nazir2017c}.

%

Let $m_{j,k}$ denote the locked-out (off)/ not-locked out
(controllable) operating state. Then,
\begin{equation}\label{eq:lockin}
m_{j,k+1} =
\begin{cases}
0, & \text{if} \quad e_{j,k} \ge e^{\textrm{max}}_{j},\\
1, & \text{if} \quad e_{j,k} < e^{\textrm{set}}_{j},\\
m_{j,k}, & \text{otherwise},
\end{cases}
\end{equation}
which implies once the SOC reaches $e^{\textrm{max}}_{j}$, the DER
enters a locked out mode until its SOC drops below a user-defined
level, $e^{\textrm{set}}_{j}$. Thus, the modified generalized battery
equation becomes,
\begin{equation}\label{SimpleModeleq1}
e_{j,k+1} = a_j e_{j,k} + \gamma_j v_{j,k} m_{j,k}. 
\end{equation}
It follows that each DER has the three operating states given in
Table~\ref{tab:op}.
The effect of adding the lockout mode is shown in
Fig.~\ref{fig:lockout}. Once reaching $e_{j,k}$ =
$e^{\textrm{max}}_{j}$ = 1 (full capacity), the DER remains off until
its SOC falls below $e_{j,k} = 0.7$. Lockout conditions associated
with $e^{\textrm{min}}_{j}$ can be modeled in a similar way.

\begin{table}[t]
	\caption{DER operating states.} \label{tab:op}
	\centering
	\begin{tabular}{|c|c|c|l|} \hline
		Op.~state & $m_{j,k}$ & $v_{j,k}$ & Outcome \\ \hline
		1 & 1 & 1 & Controllable and cleared \\ 
		2 & 1 & 0 & Controllable, but not cleared \\
		3 & 0 & 0 & Locked and turned OFF \\ \hline
	\end{tabular}
\end{table}

\subsection{Synchronization and Oscillations in TCL Simulation}

\begin{figure}[t!]
	\begin{center}
		\includegraphics[scale=0.39]{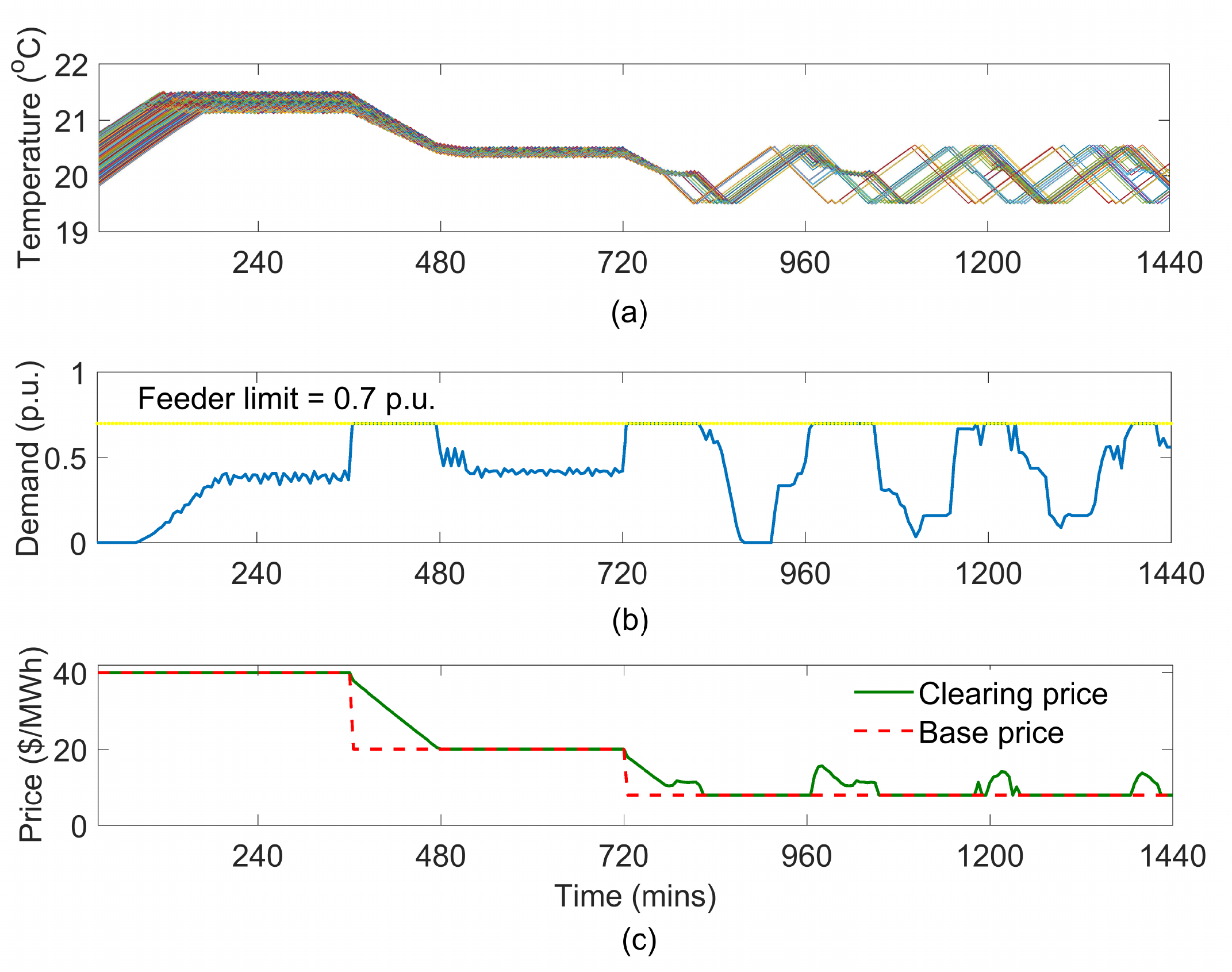}
		\caption{(a) Temperature evolution of individual TCLs, (b) 5-minute
			average aggregate demand, (c) base price and clearing price \cite{Nazir2017c}.}
		\label{fig:ncase8}
	\end{center}
\end{figure}

Equations \eqref{eq:GBM1}-\eqref{SimpleModeleq1} governing DER
operation were simulated for a population of (air conditioning) TCLs
\cite{Nazir2017c}. As illustrated in Fig.~\ref{fig:ncase8}, the TCLs
started with diverse initial temperatures. Because the base price
remained high for a few hours, most of the TCLs did not consume power
initially. Note that their temperatures became nearly synchronized at
around 200~min. Later, as the base price dropped to 20~\$/MWh, almost
all TCLs turned on. However, since the aggregate demand then exceeded
$d^{\textrm{Feeder}}$ (=~0.7~p.u.), the clearing price
increased. During minutes 480-720, the demand stayed flat. At 720~min,
when the price dropped further, the TCLs found the low price even more
favorable and consumed power. For TCLs reaching their minimum
temperature, lockout conditions were activated resulting in prolonged
on/off periods. Thus, large oscillations in the aggregate demand can
be observed. A simplified aggregate model is presented next to provide
further intuition.

\subsection{Simplified Aggregate Model}\label{sec:simpleAggregate}


When $u_k$ is relaxed to be a continuous variable, $ 0 \le u_k \le 1$,
\eqref{eq:GBM1} can express the average dynamics of a homogeneous
population of devices \cite{Zhao2017}. In the context of a large
aggregate storage device, $u_k$ represents the normalized power
consumption at time $k$ \cite{Kizilkale2014b}. Alternatively, $u_k$
can be interpreted as the probability of being ON (charging). To
establish an initial (simple) model, assume there is a proportional
relationship between the bid price and the charging decision
(i.e.~higher bids provide a higher probability of being
charged). Then,
\begin{equation}\label{SimpleModeleq2}
u_k = K_{\textrm{p}} \pi_k,
\end{equation}
where $K_{\textrm{p}} \ge 0$ is a proportional constant, which
implicitly reflects the DER's bid strength relative to the market
clearing price. Let $K_{\textrm{p}} = 1/\pi^{\textrm{max}}$. Then
$\pi_k = \pi^{\textrm{max}}$ will give $u_k = 1$.
Substituting \eqref{eq:SOCvsBid} into \eqref{SimpleModeleq2} and then
using \eqref{eq:GBM1} gives,
\begin{equation}\label{SimpleModeleq4}
u_{k+1} = (a - \gamma  \beta K_{\textrm{p}}) u_k + \pi^{\textrm{max}} K_{\textrm{p}} (1-a).
\end{equation}
Define $K_{\textrm{c}} = \pi^{\textrm{max}} K_{\textrm{p}} (1-a)$ and
$\alpha = (a - \gamma \beta K_{\textrm{p}})$. Then,
\begin{equation}
u_{k+1} = \alpha u_k + K_{\textrm{c}}, \label{SimpleModeleq6}
\end{equation}
which is a simple first order linear difference equation whose
behavior is governed by underlying parameters $\gamma$, $\beta$ and
$K_{\textrm{p}}$. The solution to \eqref{SimpleModeleq6} can be
written as,
\begin{align}\label{SimpleModelSol}
u_{k} &= \alpha^k u_{0} + K_{\textrm{c}} \sum_{j =0}^{k-1} \alpha^j ,
\end{align}
with equilibrium given by, 
\begin{align}\label{SimpleModelEQ}
u^{*} = \frac{K_{\textrm{c}}}{1 - \alpha}, \quad
e^{*} = \frac{\gamma \pi^{\textrm{max}} K_{\textrm{p}}}{1 - \alpha}, \quad
\pi^{*} = \frac{\pi^{\textrm{max}} (1-a) }{1 - \alpha} ,
\end{align}
where $\gamma,\beta,K_{\textrm{p}} \ge 0$ and $\alpha \le
a$. Moreover, since $a < 1$ for any lossy battery, $\alpha <
1$. Stability conditions can easily be derived from
\eqref{SimpleModeleq6}. When $|\alpha| <1$, the solution converges to
the equilibrium $u^*$ (i.e.~the equilibrium is stable). When $0 <
\alpha < a$ the solution is monotonic, whereas when $-1 < \alpha < 0$
the solution oscillates, with decreasing amplitude. Finally, when
$\alpha < -1$, the solution oscillates with increasing amplitude,
resulting in instability. This may occur with small values of $a$ or
with large values of $\gamma$, $\beta$ or $K_{\textrm{p}}$. Thus, the
bid curve slope $\beta$ (also mentioned in \cite{Zhao2017}), $a$,
$\gamma$ and $K_{\textrm{p}}$ all influence stability conditions.

For a numerical example, consider two cases:
\begin{itemize}
\item[(i)] $a$ = 0.9, $\gamma = 0.1$, $\beta = 50$,
  $\pi^{\textrm{max}} = 50$, $K_{\textrm{p}} = 0.02$, and
\item[(ii)] $a = 0.7$, $\gamma = 0.25$, $\beta = 150$, $\pi^{\textrm{max}} =
150$, $K_{\textrm{p}} = 0.05$.
\end{itemize}
For the first case, $\alpha = 0.8$ and the solution converges to
$u^{*}= 0.5$, $e^{*} = 0.5$ and $\pi^{*} = 25$. In the second case,
$\alpha = -1.175$ and the solution oscillates with increasing
amplitude, ultimately diverging. This suggests that a collection of
highly lossy DERs which have fast charging rates (e.g.~poorly
insulated houses with large AC units) and aggressive bid functions can
negatively impact system stability.


Thus, \eqref{SimpleModeleq6} provides valuable insights into the
response of a collection of homogeneous devices under transactive
coordination. However, it assumes that $u_k$ is a continuous variable,
$ 0 \le u_k \le 1$, whereas the on/off decisions for individual DERs
are discrete. Moreover, the effects of lockouts and feeder limits
cannot be captured using \eqref{SimpleModeleq6}.

\subsection{Influence of Feeder Limit Constraints}

Fig.~\ref{fig:ncase8} provides an example where the aggregate response
of a group of DERs exhibits oscillatory behavior following
synchronizing events. Under normal circumstances, each DER would turn
on/off periodically during periods when the base price was low. Across
a population of DERs, mixing of different frequency oscillations gives
rise to damped oscillations and beating in the aggregate demand
\cite{Docimo2017, Nazir2017c, Nazir2017a}.

However, imposing a feeder capacity limit $d^{\textrm{Feeder}}$, as in
Fig.~\ref{fig:ncase8}, can affect the aggregate DER dynamics
\cite{Nazir2017c}. A drop in market price may lead to many DERs
intending to charge simultaneously. When a strict feeder limit is
imposed, only the DERs with sufficiently high bids will initially be
cleared. DERs with relatively lower bids wait until their bids rise
sufficiently for them to be cleared. As a result, the DERs' on/off
periods are affected by the feeder limit. Such a constraint can
induce oscillations and beating, and may also lead to limit cycle
behavior \cite{Pikovsky2003, Docimo2017, Hiskens2007}.

While transactive coordination can be influenced by a variety of
phenomena (e.g.~hysteresis, lockout, base price behavior and the
feeder limit), a unified framework for studying such effects is
lacking. Parametric studies using individual dynamic equations can be
useful but cumbersome. Hence, efficient techniques for analyzing
aggregate behavior are required. Economic and stable operation of the
overall power system can be ensured by developing representative DER
aggregate models and incorporating them into appropriate multi-period
optimization problems.


\section{Aggregate Model under Market-based Coordination of DERs} \label{sec:agg}

It has been shown in \cite{Koch2011, Bashash2011, Mathieu2013,
  Zhang2013, Totu2016, Nazir2016} that the natural dynamics of TCLs
can be expressed in a bin model structure where each bin represents a
temperature range and an on/off state. The evolution of the
probabilities of TCLs lying in each bin can be described by,
\begin{equation}
X_{k+1} = A X_{k}, \label{eq:AX}
\end{equation}
where $A$ is the transpose of the Markov transition matrix that is
constructed from the probabilities associated with transitions between
bins. It has been shown that the Markov transition matrix can be
obtained analytically from the difference equations governing the TCL
thermal dynamics \cite{Mathieu2013, Lygeros2015, Nazir2017a,
  Nazir2016}, or by applying system identification techniques
\cite{Koch2011, Mathieu2013}.

Existing bin-based aggregate models, however, do not capture the
influence of transactive coordination on the behavior of DERs. There
has also been limited attention given to synchronization
issues. Hence, an aggregate model that incorporates price will now be
developed. Lockout constraints and other mechanisms that eliminate
synchronization will also be incorporated. This bin model is then used
within an MPC framework in Section~\ref{sec:MPC}.

\subsection{Homogeneous Population Model}\label{sec:HomBinModel}

The aggregate bin model expressing the dynamics of DERs (e.g.~TCLs,
storage devices, electric vehicles) under transactive coordination
depends both on SOC dynamic equations and user bid curves. For a
homogeneous population, the storage dynamic equation \eqref{eq:GBM1}
becomes,
\begin{equation}\label{eq:GBM10}
e_{j,k+1} = a e_{j,k} + \gamma u_{j,k} , 
\end{equation}
and the relationship \eqref{eq:SOCvsBid} between bid price and SOC is,
\begin{equation}\label{BinModeleq20}
\pi_{j,k} = \pi^{\textrm{max}} - \beta e_{j,k}, 
\end{equation}
where the maximum bid is $\pi^{\textrm{max}}$ when $e_{j,k} = 0$ and
the minimum bid is $\pi^{\textrm{min}} = \pi^{\textrm{max}} - \beta$
when $e_{j,k} = 1$. It is assumed that when $e_{j,k}$ reaches 1, the
DER is automatically switched off.

Assume market clearing occurs every $\tau$ minutes, during which it is
reasonable to assume there would be discrete shifts in the SOC levels
and the price bids. Consider $N_{\textrm{B}}$ bins (i.e.~discrete
price intervals) between $\pi^{\textrm{max}}$ and
$\pi^{\textrm{min}}$, with each bin's width being
$(\pi^{\textrm{max}}- \pi^{\textrm{min}})/N_{\textrm{B}}$. The bins
have indices $i = 1,...,N_{\textrm{B}}$. The bins are organized by
decreasing price levels, with $\tilde{\pi}_0 = \pi^{\textrm{max}}$ and
$\tilde{\pi}_{N_{\textrm{B}}} = \pi^{\textrm{min}}$. The $i$-th bin's
boundaries are $\{\tilde{\pi}_{i-1},\tilde{\pi}_i\}$. For a
homogeneous model, a direct mapping to the discrete SOC levels can be
obtained. Let $\tilde{e}_i$, $i = 0,1,...,N_{\textrm{B}}$, be the
discrete SOC levels associated with $\tilde{\pi}_i$, so that
$\tilde{e}_0 = e^{\textrm{min}} = 0$ and $\tilde{e}_{N_{\textrm{B}}} =
{e}^{\textrm{max}} = 1$.
\begin{figure}[t!] 
\begin{center}
\includegraphics[scale=0.25]{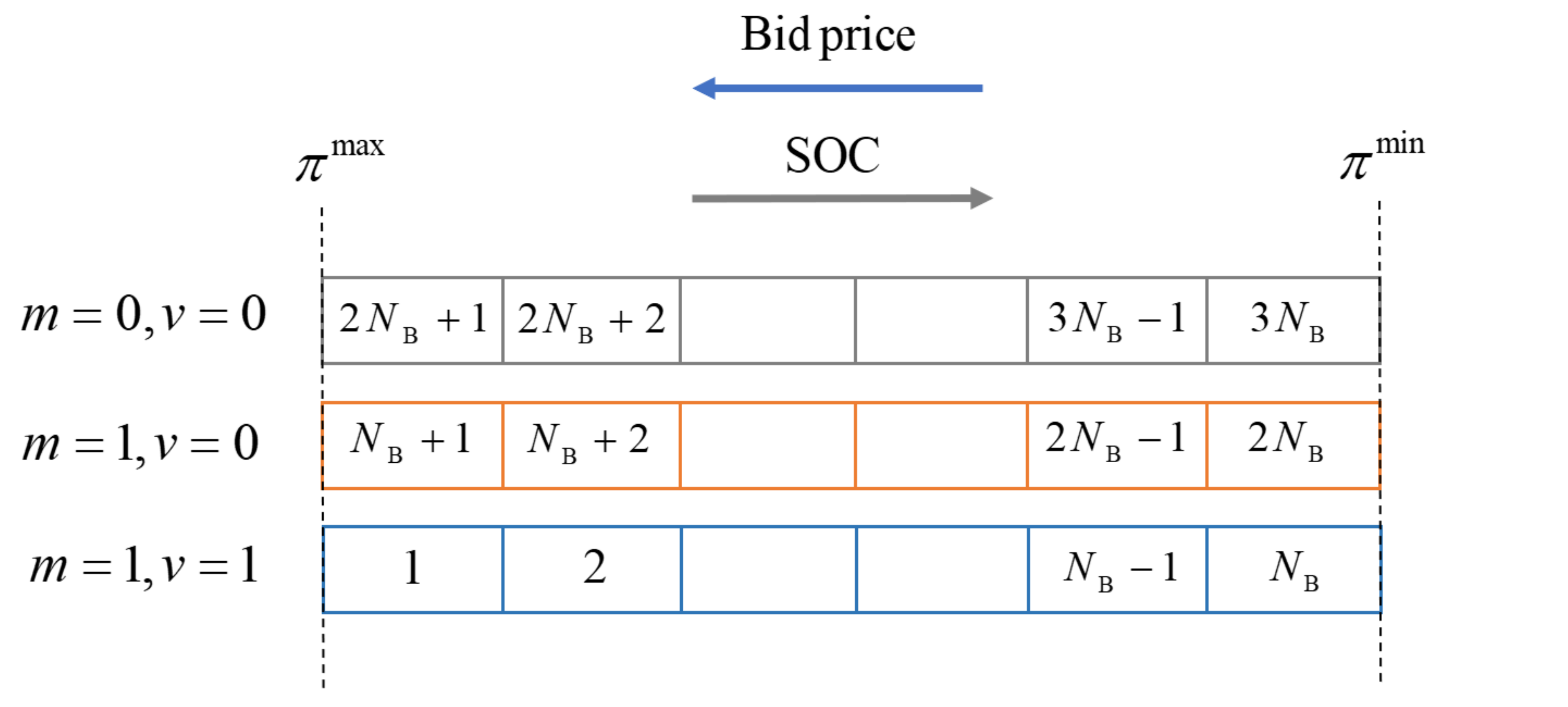}
\caption{Bin model for a homogeneous population.  Bid prices decrease
  from left to right whereas SOC levels increase. The three different
  operating states are marked on the left.} \label{fig:Bin3stageHom}
\end{center}
\end{figure}

Note that each bin also needs to consider the on/off and locked status
outlined in Table~\ref{tab:op}. Hence, a three-stage bin model is
proposed, as illustrated in Fig.~\ref{fig:Bin3stageHom}, where bins
are mapped to the operating states by defining the sets,
\begin{subequations}
\begin{alignat}{2}
I_1 &= \{1,...,N_{\textrm{B}}\} &&\text{ for } m=1, v=1 \\
I_2 &= \{N_{\textrm{B}}+1,...,2N_{\textrm{B}}\} &&\text{ for } m=1, v=0 \\
I_3 &= \{2N_{\textrm{B}}+1,...,3N_{\textrm{B}}\} &&\text{ for } m=0, v=0.
\end{alignat}
\end{subequations}
Let $x_{i,k} \ge 0$ be the fraction of DERs lying inside bin $i$ at
time $k$, where $i \in I_1$, $I_2$ or $I_3$, depending on the bid
price (or the SOC level) and the operating state. Let $p_{j,i}$ be the
probability of transitioning from bin $j$ to bin $i$ in one time
step. Then the state transitions for the $i$-th bin can be written,
\begin{equation}
x_{i,k+1} = \sum_{j=1}^{3N_{\textrm{B}}} p_{j,i} x_{j,k}, \quad \forall i, k.
\end{equation}
This can be expressed in matrix form \eqref{eq:AX} as,
\begin{equation}\label{(Amatrix)}
A =
\begin{bmatrix}
p_{1,1} & p_{1,2} & \cdots & p_{1,3N_{\textrm{B}}} \\
p_{2,1} & p_{2,2} & \cdots & p_{2,3N_{\textrm{B}}} \\
\vdots  & \vdots  & \ddots & \vdots  \\
p_{3N_{\textrm{B}},1} & p_{3N_{\textrm{B}},2} & \cdots & p_{3N_{\textrm{B}},3N_{\textrm{B}}} 
\end{bmatrix}^T.
\end{equation}
Note that $0 \le p_{j,i} \le 1, \; \forall i,j$, and
$\sum_{i=1}^{3N_{\textrm{B}}} p_{j,i}=1, \; \forall j$.

Assume the clearing price is $\pi^{\textrm{clr}}_k$. Then, all TCLs in
bins with bid prices greater than $\pi^{\textrm{clr}}_k$ will be
cleared. For a fixed value of $\pi^{\textrm{clr}}_k$, the transition
probabilities $p_{j,i}$ can be estimated from a large number of
samples by observing the evolution of DERs after the market clears at
$\tau$~minutes.

The clearing price $\pi^{\textrm{clr}}_k$ determines which bins are
cleared. Hence, the $p_{j,i}$ are functions of $\pi^{\textrm{clr}}_k$
and the $A$-matrix is time-varying for varying
$\pi^{\textrm{clr}}_k$. The aggregate model can be expressed as,
\begin{equation}\label{eq:HybridModel}
X_{k+1} = A_k X_{k}. 
\end{equation}
If $\pi^{\textrm{clr}}_k$ remains constant then $A_k = A, \forall
k$. Section~\ref{sec:ModelPerformance} illustrates the construction of
the $A$-matrices, the connection between system behavior and the
eigenvalues of $A_k$, and model performance.

\subsection{Bin Model under Parameter Heterogeneity}\label{sec:HeteroBinModel}

When parameters are heterogeneous, bins are defined similarly to
Section~\ref{sec:HomBinModel}, except $\tilde{\pi}_0 = \text{max
}\pi^{\textrm{max}}_j$ and $\tilde{\pi}_{N_{\textrm{B}}} = \text{min
}\pi^{\textrm{min}}_j$. Compared to the homogeneous case, however, the
discretization in SOC levels cannot be mapped directly to the bid
price levels, unless additional bins are added to track both SOC
levels and bid prices. The $A$-matrix elements can be obtained from a
large number of samples. Alternatively, the dynamics of the
heterogeneous population can be captured by considering clusters of
homogeneous groups and their respective transition equations
\cite{Nazir2017a, Alizadeh2015}.  

\subsection{Reformulation of the Transition
  Equations} \label{sec:TransitionEquations}

While \eqref{eq:HybridModel} can be used to simulate the aggregate
behavior of DERs, the price signal must be known {\em a priori}. To
determine the optimal price signal and resultant cleared demand in a
multi-period optimization problem, the influence of price variation
must be separated from the propagation dynamics
\eqref{eq:AX}. This can be acomplished by decomposing the state
associated with each bin into ON and OFF fractions,
\begin{equation}\label{eq:trans1}
x_{i,k} = x^{\textrm{on}}_{i,k} + x^{\textrm{off}}_{i,k}, \quad
x^{\textrm{on}}_{i,k}, x^{\textrm{off}}_{i,k} \ge 0, \quad \forall
i \in I_1 \cup I_2, \forall k,
\end{equation}
where $x^{\textrm{on}}_{i,k}$ and $x^{\textrm{off}}_{i,k}$ are
decision variables determined by the optimization process. More
specifically, the division between ON and OFF fractions of a bin is
determined by a comparison between the bid price associated with that
bin and the clearing price $\pi^{\textrm{clr}}$.

Given values for the decision variables $x^{\textrm{on}}_{i,k}$ and
$x^{\textrm{off}}_{i,k}$, the states are reset according to,
\begin{subequations}
\begin{align}
x_{i,k}^+ &= x^{\textrm{on}}_{i,k} + x^{\textrm{on}}_{i+N_B,k}, \quad
\forall i\in I_1, \forall k, \\
x_{i,k}^+ &= x^{\textrm{off}}_{i,k} + x^{\textrm{off}}_{i-N_B,k}, \quad
\forall i\in I_2, \forall k,
\end{align}
\end{subequations}
where the `+' superscript indicates reset values. Note also that for
the locked bins, $i \in I_3$, no DERs can be turned on. Hence,
\begin{equation}\label{eq:trans2}
x_{i,k}^+ = x_{i,k} = x^{\textrm{off}}_{i,k},  \quad \forall i \in I_3, \; \forall k.
\end{equation}
These reset equations can be expressed in matrix form,
\begin{equation}
X_k^+ = B^{\textrm{on}} X_k^{\textrm{on}} + B^{\textrm{off}}
X_k^{\textrm{off}}. \label{eq:reset}
\end{equation}
Starting from the reset state values, the evolution of the bin
probabilities has a similar form to \eqref{eq:AX},
\begin{equation}
X_{k+1} = A X_k^+, \label{eq:resetprop}
\end{equation}
where $A$ now captures the natural dynamics of DERs, as expressed by
\eqref{eq:GBM10} and \eqref{BinModeleq20}.

Finally, it should be noted that the structure of $A$ ensures
conservation of probability,
\begin{equation}\label{eq:trans5}
\sum_{i=1}^{3N_{\textrm{B}}} x_{i,k} = 1, \quad \forall k, 
\end{equation}
and $x_{i,k} \ge 0, \forall k$. Note that with this reformulation
all equations remain linear.  This is especially helpful for MPC
design, as shown in Section~\ref{sec:MPC}.

\begin{figure}[t!]
	\begin{center}
		\includegraphics[scale=0.41]{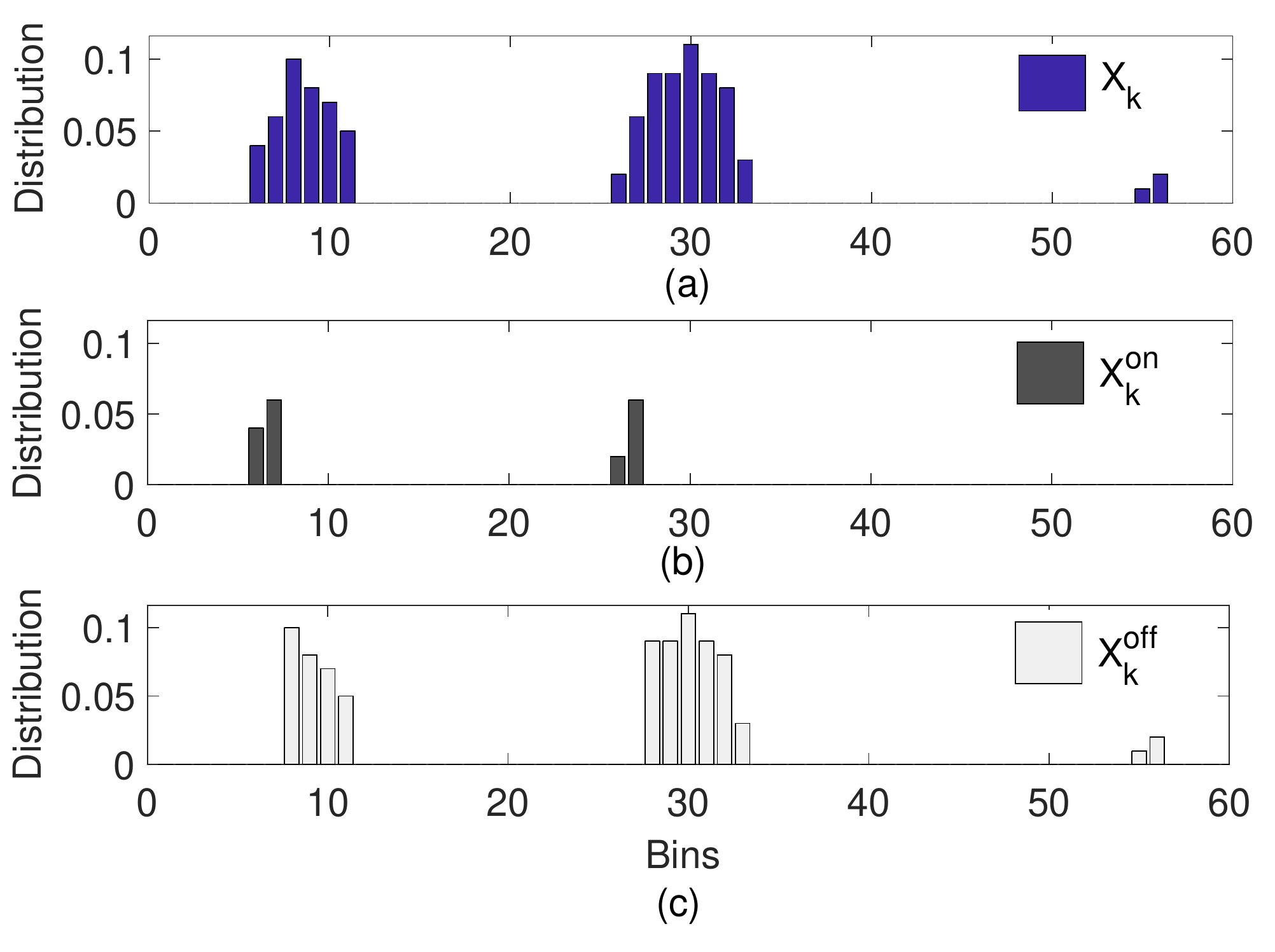}
		\caption{Distributions for (a) $X_k$, (b) $X^{\textrm{on}}_{k}$ and (c) $X^{\textrm{off}}_{k}$. }
		\label{fig:PMFfig}
	\end{center}	
\end{figure}

An example with $N_{\textrm{B}} = 20$ (total 60 bins) is shown in
Fig.~\ref{fig:PMFfig}. Bins lie in three sets $I_1 = \{1,...,20\}$,
$I_2 = \{21,...,40\}$, and $I_3 = \{41,...,60\}$. The overall
distribution, $X_k$ at time $k$, is shown in
Fig.~\ref{fig:PMFfig}(a). Recall from the bin definitions and from
Fig.~\ref{fig:Bin3stageHom} that bins from left to right have
decreasing bid prices. Also, while $I_1$, $I_2$, $I_3$ differ in their
operating states, the bid prices at bins $i$, $i+N_{\textrm{B}}$ and
$i+2N_{\textrm{B}}$, where $i = 1,...,20$, are the same. Assume,
$\pi^{\textrm{clr}}_k = \tilde{\pi}_7$. Thus, all DERs in bins 6, 7,
26 and 27 are cleared. The ON and OFF distributions,
$X^{\textrm{on}}_{k}, X^{\textrm{off}}_{k}$, are shown in
Figs.~\ref{fig:PMFfig}(b) and~\ref{fig:PMFfig}(c). By
\eqref{eq:reset},\eqref{eq:resetprop}, $B^{\textrm{on}}$ and
$B^{\textrm{off}}$ will then act on $X^{\textrm{on}}_{k}$ and
$X^{\textrm{off}}_{k}$, respectively, to give $X_{k+1}$.

\subsection{Logic Equations for Market Clearing}

An arbitrary optimizer could choose the ON/OFF quantities from each
bin as long as \eqref{eq:trans1}-\eqref{eq:trans5} were
satisfied. However, in a transactive dispatch mechanism, DERs
with higher bid prices than the clearing price are cleared. Hence,
additional logic is required to simulate behavior under the
transactive market clearing mechanism. For example, with the bin
definitions of Fig.~\ref{fig:Bin3stageHom}, bins from right to left
have increasing bid prices, and so should have higher priority to turn
on. To accomplish this, binary variables can be used.

Let $u^{\textrm{on}}_{i,k} \in \{0,1\}$ be the binary on/off decision
associated with clearing all TCLs in bin $i$, and consider the
following set of equations:
\begin{subequations}
\begin{align}
0 \le x^{\textrm{on}}_{i,k} \le u^{\textrm{on}}_{i,k}, &\quad i = 1,...,2N_{\textrm{B}} \label{eq:nonlin1a} \\
0 \le x^{\textrm{off}}_{i,k} \le u^{\textrm{off}}_{i,k}, &\quad i = 1,...,3N_{\textrm{B}} \label{eq:nonlin1b} \\
u^{\textrm{on}}_{i,k} + u^{\textrm{off}}_{i,k} = 1, &\quad i = 1,...,3N_{\textrm{B}} \label{eq:nonlin1c} \\
u^{\textrm{on}}_{i,k} \ge u^{\textrm{on}}_{i+1,k} , &\quad i =
	1,..., N_{\textrm{B}}-1  \label{eq:nonlin1d}	\\
u^{\textrm{on}}_{i,k} = u^{\textrm{on}}_{i+N_{\textrm{B}},k}, &\quad i = 1,...,N_{\textrm{B}}.  \label{eq:nonlin1f} 	
\end{align}
\end{subequations}
If $x^{\textrm{on}}_{i,k}$ DERs from bin $i$ are to be chosen to be
ON, we need $u^{\textrm{on}}_{i,k}$ = 1, otherwise
$u^{\textrm{on}}_{i,k}$ = 0. This is accomplished by
\eqref{eq:nonlin1a}. Likewise, by \eqref{eq:nonlin1b},
$u^{\textrm{off}}_{i,k} \in \{0,1\}$ is used for choosing
$x^{\textrm{off}}_{i,k}$. Ensuring that each bin has only ON or OFF
DERs is achieved by \eqref{eq:nonlin1c}. (We assume that when a bin is
cleared, all DERs in that bin turn on.) Next, \eqref{eq:nonlin1d}
ensures that bins with higher bid prices must be turned ON before bins
with lower prices can be chosen. Finally, \eqref{eq:nonlin1f} ensures
that bins with the same bid prices, but different operating states
$I_1$ and $I_2$ get cleared simultaneously. Note that the logic
equations \eqref{eq:nonlin1a}-\eqref{eq:nonlin1f} are linear in
$x^{\textrm{on}}_{i,k}$, $x^{\textrm{off}}_{i,k}$,
$u^{\textrm{on}}_{i,k}$, and $u^{\textrm{off}}_{i,k}$.


\section{Model Predictive Control Formulation}\label{sec:MPC}


The bin-based aggregate model of DERs can be incorporated into a model
predictive control framework to determine the DER schedule that gives
minimum power supply cost over a finite horizon. Let the distribution
network be supplied by power sources $P_k^s$, $s=1,...,N_\textrm{s}$ at time
$k$. Each source has cost $C_s(P_k^s)$ which is typically
quadratic. The overall demand of the network at time $k$ is $D_k$,
which is composed of controllable DER demand $D_k^c$ and
uncontrollable demand $D_k^o$ of other loads. The cost minimization
problem can be formulated as,
\begin{subequations}
\begin{alignat}{2}
\min \quad &\sum_{k=1}^{N_\textrm{k}} \sum_{s=1}^{N_\textrm{s}} C_s(P^s_{k})  \label{eq:obj} \\
\text{s.t.}  \quad \sum_{s=1}^{N_\textrm{s}} P^s_k &= D_k , &
\quad &\forall  k \label{eq:pbalance} \\
D_k &= D^c_k + D^o_k,  &  \quad &\forall  k,    \label{eq:dbalance} \\
D^c_k &= \sum_{i=1}^{3N_{\textrm{B}}}
x^{\textrm{on}}_{i,k}, & \quad &\forall k, \label{eq:bintrans2d} \\
x_{i, k=1}  &= x^{\textrm{ini}}_i, & \quad &\forall i, \label{eq:bintrans2f}\\
D_k &\le d^{\textrm{Feeder}} &  \quad &\forall k,  \label{eq:FeederLimit}
\end{alignat}
\end{subequations}
along with the transition equations
\eqref{eq:trans1}-\eqref{eq:trans5} and logic constraints
\eqref{eq:nonlin1a}-\eqref{eq:nonlin1f}.

The objective function \eqref{eq:obj} minimizes the total cost of
supply. Supply-demand balance is enforced by \eqref{eq:pbalance}. The
Lagrange multiplier associated with this constraint represents the
electricity price, $\lambda^{\textrm{elec}}_k$. Total demand is
established by \eqref{eq:dbalance}, with \eqref{eq:bintrans2d}
relating aggregate controllable load to the ON fractions of all
bins. Initial conditions are established by \eqref{eq:bintrans2f},
where $X^{\textrm{ini}}$ is assumed to be known. Finally,
\eqref{eq:FeederLimit} ensures the feeder limit is not
violated. Individual capacity and ramp limits for each supplier may
also be incorporated.

Additionally, to avoid arbitrarily setting $u^{\textrm{on}}_{i,k} =
1$, a cost $\tilde{C}_i(u^{\textrm{on}}_{i,k})$ is included in the objective,
\begin{equation}\label{eq:nonlinObj} 
\sum_{k=1}^{N_\textrm{k}} \sum_{s=1}^{N_\textrm{s}} C_s(P^{\textrm{s}}_k) + \mu_{\textrm{w}} \sum_{k=1}^{N_\textrm{k}} \sum_{i=1}^{3N_{\textrm{B}}} \tilde{C}_i(u^{\textrm{on}}_{i,k}),
\end{equation}
where $\mu_{\textrm{w}}$ is a tuning parameter. Note,
$\tilde{C}_i(u^{\textrm{on}}_{i,k}) = 1, \forall i,k$ has been used in
this paper.

Due to the presence of both continuous and binary variables, the
overall formulation is a mixed integer programming (MIP) problem. A
mixed integer linear program (MILP) can be obtained by approximating
the suppliers' quadratic cost functions with piecewise linear
segments.

\subsection{Measures to Avoid DER Synchronization}

Synchronization occurs when a large fraction of DERs lie within a
narrow range of bins. It may lead to large oscillations in the
aggregate demand and high volatility in clearing prices. Hence, to
avoid this situation, let $b^{\textrm{max}}$ be the maximum allowable
fraction in any bin, and enforce the constraint,
\begin{equation}
x_{i,k} \le b^{\textrm{max}}, \quad \forall i,k.
\end{equation}
Alternatively, a quadratic cost term can be added in the objective
function,
\begin{equation} \label{eq:obj3}  
\sum_{k=1}^{N_\textrm{k}} \sum_{s=1}^{N_\textrm{s}} C_s(P^{\textrm{s}}_k) + \mu_{\textrm{w}} \sum_{k=1}^{N_\textrm{k}} \sum_{i=1}^{3N_{\textrm{B}}} \tilde{C}_i(u^{\textrm{on}}_{i,k}) + \mu_{\textrm{s}} \sum_{i=1}^{3N_{\textrm{B}}} (x_{i,k} - b^{\textrm{avg}})^2, 
\end{equation}
where $b^{\textrm{avg}} = 1/(3N_{\textrm{B}})$ and $\mu_{\textrm{s}}$
can be adjusted to distribute DERs widely over the bin space.

\subsection{QP Formulation}

Instead of using binary variables in
\eqref{eq:nonlin1a}-\eqref{eq:nonlin1f} and \eqref{eq:nonlinObj}, a
relaxed QP formulation can be established by assigning costs per bin,
$\hat{C}_i(x^\textrm{on}_{i,k})$, that increase with increasing bin
index. For example, $\hat{C}_i(x^\textrm{on}_{i,k})= w_o^{i}
x^\textrm{on}_{i,k}, \forall i \in I_1$ and
$\hat{C}_i(x^\textrm{on}_{i,k}) = w_o^{i- N_{\textrm{B}}}
x^\textrm{on}_{i,k}, \forall i \in I_2$, where $w_o > 1$. (The
simulations in Section~\ref{sec:Results} use $w_o = 3$.) The modified
objective function becomes,
\begin{equation}  \label{eq:obj5} 
\sum_{k=1}^{N_\textrm{k}} \sum_{s=1}^{N_\textrm{s}} C_s(P^{\textrm{s}}_k) + \mu_{\textrm{w}} \sum_{k=1}^{N_\textrm{k}} \sum_{i=1}^{3N_{\textrm{B}}} \hat{C}_i(x^{\textrm{on}}_{i,k}) + \mu_{\textrm{s}} \sum_{i=1}^{3N_{\textrm{B}}} (x_{i,k} - b^{\textrm{avg}})^2, 
\end{equation}
where $\mu_{\textrm{w}}$ is a tuning parameter. 

The overall formulation consists of the objective \eqref{eq:obj5},
constraints \eqref{eq:pbalance}-\eqref{eq:FeederLimit} and transition
equations \eqref{eq:trans1}-\eqref{eq:trans5}. All constraints become
linear and the objective function remains quadratic. Overall, we
obtain an efficient QP form.

\subsection{Transactive Dispatch Rule}

Once the above problem has been solved, the indices of the ON bins can
be recovered from $x^{\textrm{on}}_{i,k}$ or
$u^{\textrm{on}}_{i,k}$. Define $i^{\textrm{max}}_k$ as the largest
index among the ON bins at period $k$ (for $i \in I_2$ subtract
$N_{\textrm{B}}$ from bin indices). For the MIP solution,
\begin{equation}  \label{eq:j0} 
i^{\textrm{max}}_k = \max_i \{u^{\textrm{on}}_{i,k} = 1\} , \quad \forall k.
\end{equation}
Since $u^{\textrm{on}}$ does not appear in the QP formulation, in that
case a bin is considered to be cleared when $x^{\textrm{on}}_{i,k}$ is
larger than a threshold $\zeta$,
\begin{equation}  \label{eq:j1} 
i^{\textrm{max}}_k = \max_i \{x^{\textrm{on}}_{i,k} \ge \zeta \} , \quad \forall k.
\end{equation}
From the SOC-bin mapping of Fig.~\ref{fig:Bin3stageHom}, the price
associated with $i^{\textrm{max}}_k$ becomes the clearing price
$\pi^{\textrm{clr}}_{k}$, that is broadcast to all the DERs for time
period $k$. Upon receiving this price, all DERs in bins $i \le
i^{\textrm{max}}_k$, (i.e.~bins with higher bid prices) should
dispatch,
\begin{equation}
u_{j,k}=
\begin{cases}
1, & \text{if} \quad \pi_{j,k} \ge \tilde{\pi}_{i^{\textrm{max}}_k},\\
0, & \text{if} \quad \pi_{j,k} < \tilde{\pi}_{i^{\textrm{max}}_k}.
\end{cases}
\end{equation}

\subsection{Practical Considerations}

Assume all DERs are managed by a single aggregator in a distribution
system. To obtain $X^{\textrm{ini}}$, which is required for
\eqref{eq:bintrans2f}, an aggregator could collect local measurements
from DERs typically near the end of an MPC horizon. Each DER's on/off,
locked status and its associated bid price are used to construct
$X^{\textrm{ini}}$. Since only aggregate information is needed, each
DER sends updates anonymously. Assuming a distribution system operator
(DSO) solves the MPC, the aggregator would send $X^{\textrm{ini}}$ to
the DSO\@. The DSO also has access to the $A$-matrix of
\eqref{eq:resetprop}, which the aggregator estimates separately. The
DSO solves the MPC and sends the aggregator the clearing prices
$\pi^{\textrm{clr}}_{k}, k = 1,...,N_\textrm{k}$. The price
$\pi^{\textrm{clr}}_{k}$ can be revealed to DERs either just for the
next market clearing interval or for several periods ahead, depending
on communication bandwidth availability. Since MPC already accounts
for the feeder limit, the two-step market clearing process described
in Section~\ref{sec:MarketBasic} is not necessary. However, if the
mechanism in Section~\ref{sec:MarketBasic} is followed, individual DER
bids additionally need to be collected at every market clearing
interval.

\section{Simulation Results}\label{sec:Results}

\subsection{Data}

\begin{figure}[t]
	\begin{center}
		\includegraphics[scale=0.34]{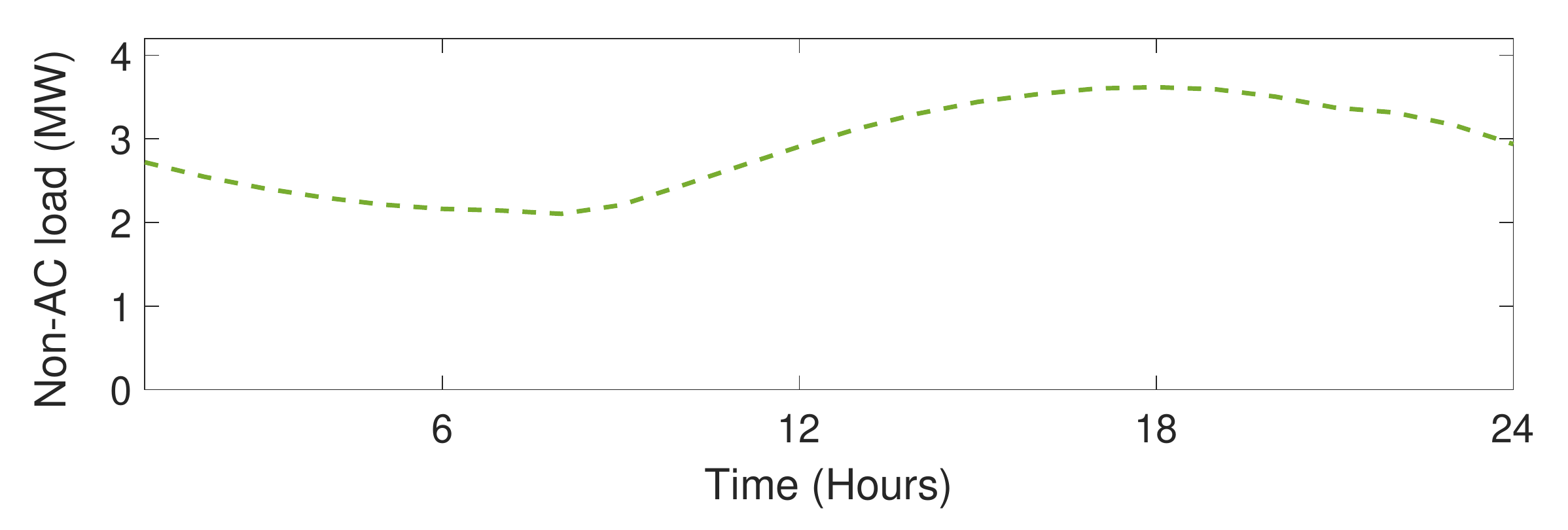}
		\caption{Aggregate demand profile excluding air-conditioner (AC) demand.}
		\label{fig:NonACdata}
	\end{center}	
\end{figure}

The case study considers a typical distribution system serving
predominantly residential loads (1,473 customers) located in Austin,
TX \cite{Dubey2015}. The system peak demand was recorded at 7.77
MW\@. Energy usage analysis of real data from 88~single-family houses
in the Mueller neighborhood of Austin, from July 2012 to June 2013,
was undertaken in \cite{Perez2014}. It was found that approximately
47\% of the peak household demand was consumed by air-conditioner (AC)
units during summer peak days. Based on this analysis and hourly
demand data from the Electric Reliability Council Of Texas (ERCOT)
\cite{ERCOT}, the non-AC demand profile of Fig.~\ref{fig:NonACdata}
was estimated. Since each AC consumes approximately 3~kW on average
(see Appendix~\ref{sec:TCLmodel}), the maximum AC load can be up to
4.4~MW, which would result in a peak demand exceeding 8~MW.

\subsection{Aggregate Model Performance}\label{sec:ModelPerformance}

To study the performance of the aggregate model
\eqref{eq:HybridModel}, first the $A$-matrix coefficients need to be
obtained. Consider a homogeneous population with $\beta_j =
40$~\$/MWh, $\pi^{\textrm{max}}_j =$ \$50, $\forall j$. With $\tau=
10$~min and known price signals, 1000~TCLs were simulated. For TCLs
originating in a specific sending bin, we can find the range of bins
reached by TCLs at the end of 10~minutes. Repeating for all bins and
normalizing these quantities, the transition probabilities (thus, the
$A$-matrix) for a known price signal were obtained.

\begin{figure}[t!]
	\begin{center}
		\includegraphics[scale=0.35]{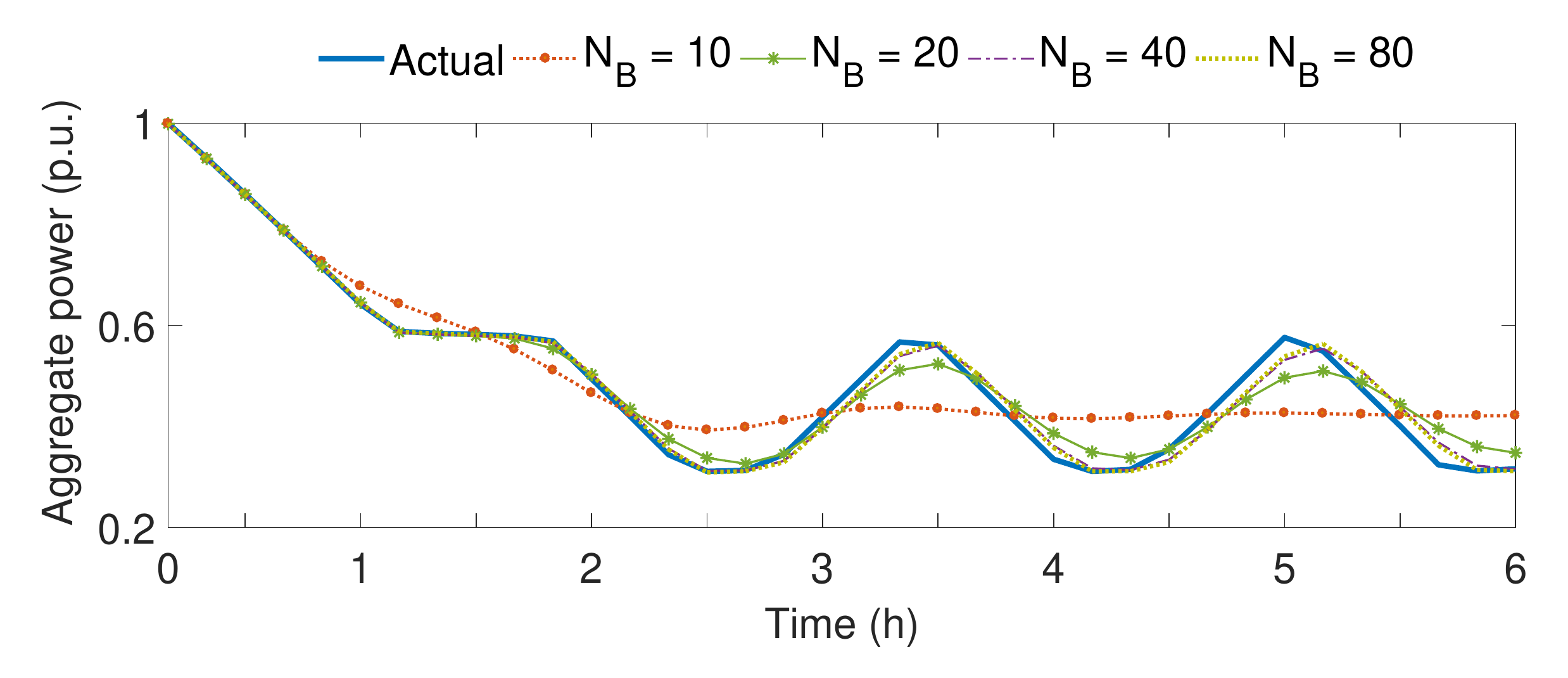}
		\caption{Aggregate demand profiles, for varying $N_{\textrm{B}}$, with clearing price at 10 $\$$/MWh (TCL initial temperatures uniform within 19 to 21$^\textrm{o}$C).}
		\label{fig:VaryNbuniforminiSOCfull}
	\end{center}	
\end{figure}
\begin{figure}[t!]
	\begin{center}
		\includegraphics[scale=0.35]{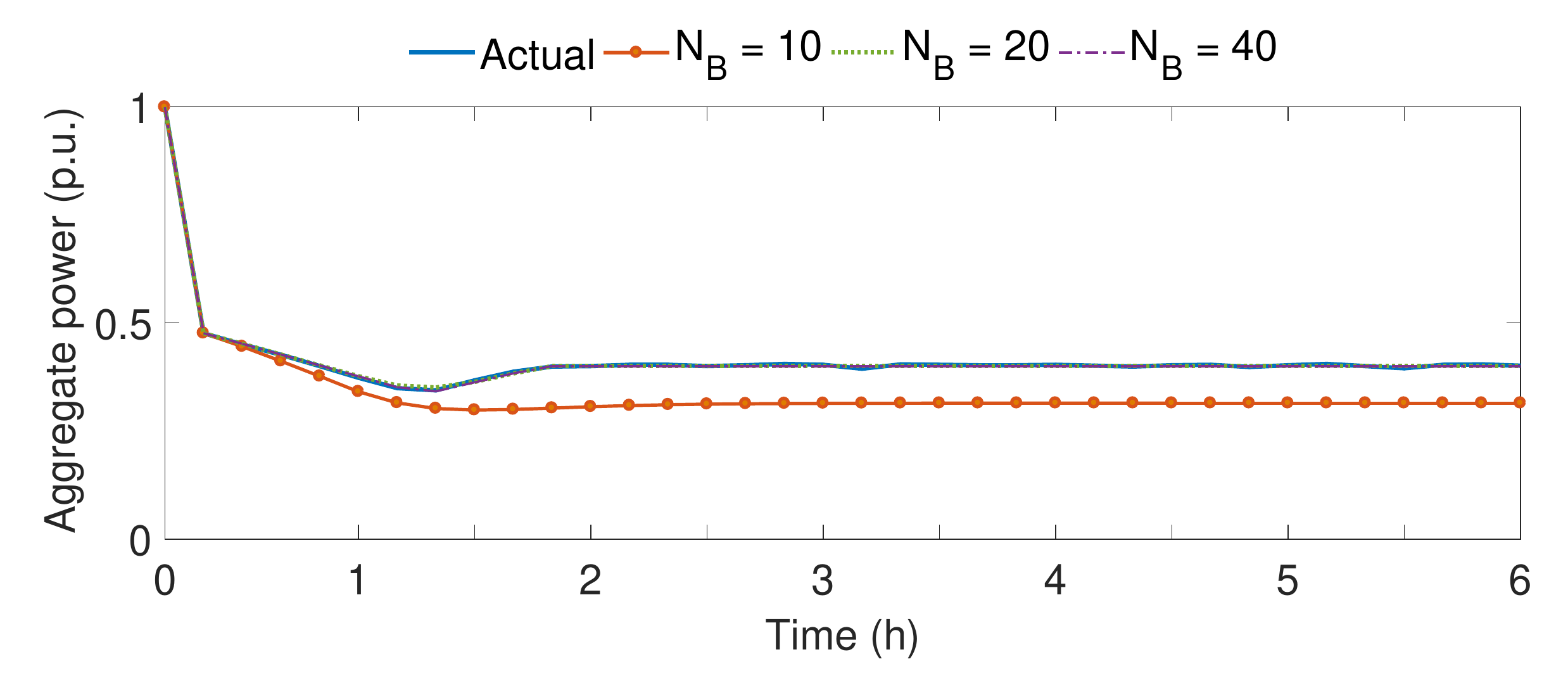}
		\caption{Aggregate demand profiles, for varying $N_{\textrm{B}}$, with clearing price at 30 $\$$/MWh (TCL initial temperatures uniform within 19 to 21$^\textrm{o}$C).}
		\label{fig:VaryNbuniforminiClearHalf}
	\end{center}	
\end{figure}

First, consider $\pi^{\textrm{clr}} = 10$ \$/MWh. All TCLs have bids
just above the clearing price, hence can get cleared. The aggregate
power consumed by 1000~TCLs over 6~hours has been shown in
Fig.~\ref{fig:VaryNbuniforminiSOCfull}. Note that the initial
temperatures of TCLs were distributed uniformly between 19 to
21$^\textrm{o}$C. With $\tau = 10$~minutes, the aggregate behavior was
also simulated using bin models of various orders, $N_{\textrm{B}} =
$10, 20, 40, 80. With $N_{\textrm{B}} = 10$ the aggregate demand
profile deviates significantly from the actual. The profiles obtained
with the other models matched the actual reasonably
well. Fig.~\ref{fig:VaryNbuniforminiClearHalf} shows the performance
for varying $N_{\textrm{B}}$ when $\pi^{\textrm{clr}} = 30$
\$/MWh. Again, the profile obtained using the $N_{\textrm{B}} = 10$
model deviates from the actual.

\begin{figure}[t!]
	\begin{center}
		\includegraphics[scale=0.347]{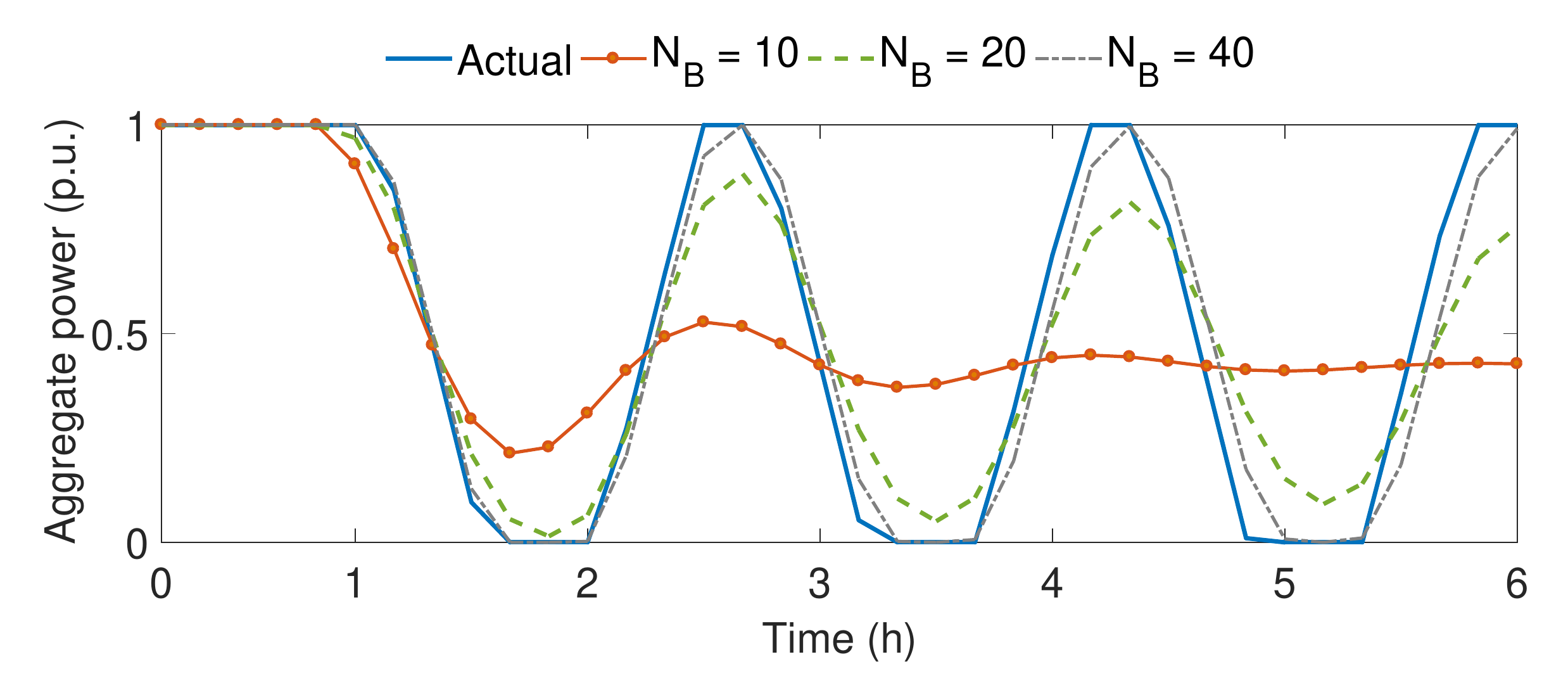}
		\caption{Aggregate demand profiles, for varying $N_{\textrm{B}}$, with clearing price at 10 $\$$/MWh (TCL initial temperatures uniform within 19.8 to 20.2$^\textrm{o}$C).}
		\label{fig:VaryNbnarrowINI_SOCfull}
	\end{center}	
\end{figure}

Next, the $A$-matrix obtained for 10~$\$$/MWh was used to simulate the
aggregate demand with initial temperatures of TCLs distributed
uniformly between 19.8 and 20.2$^\textrm{o}$C. Profiles are shown in
Fig.~\ref{fig:VaryNbnarrowINI_SOCfull}. Since most TCLs turned on/off
almost at the same time, the oscillation amplitudes were larger
compared to those in Fig.~\ref{fig:VaryNbuniforminiSOCfull}.

With the price signal fixed at $\pi^{\textrm{clr}} = 30$~\$/MWh, the
aggregate demand reaches a constant level, whereas with
$\pi^{\textrm{clr}} = 10$~\$/MWh, oscillatory behavior has been
observed. This is because with a low clearing price, TCLs enter the
locked mode when their temperatures reach 19$^\textrm{o}$C and again
become controllable when temperatures exceed
19.6$^\textrm{o}$C. Analyzing the eigenvalues of the $A$-matrices, we
observed that for the $A$-matrix with $\pi^{\textrm{clr}}$ =
30~\$/MWh, the eigenvalues have only real parts. For the $A$-matrix
with $\pi^{\textrm{clr}} = 10$~\$/MWh, pairs of complex eigenvalues
exist, suggesting an oscillatory response. Ideally under
$\pi^{\textrm{clr}} = 10$, a homogeneous population exhibits undamped
oscillations, whereas the eigenvalues of the $A$-matrix suggest damped
oscillations. This discrepancy exists due to modeling error from the
discretization \cite{Koch2011, Bashash2013, Nazir2017a}. However, as
shown in Figs.~\ref{fig:VaryNbuniforminiSOCfull}
and~\ref{fig:VaryNbnarrowINI_SOCfull}, with sufficiently large
$N_{\textrm{B}}$ the actual behavior can be tracked closely for
several hours, which is suitable for control purposes
\cite{Bashash2013, Nazir2017a}.
\begin{figure}[t!]
	\begin{center}
		\includegraphics[scale=0.345]{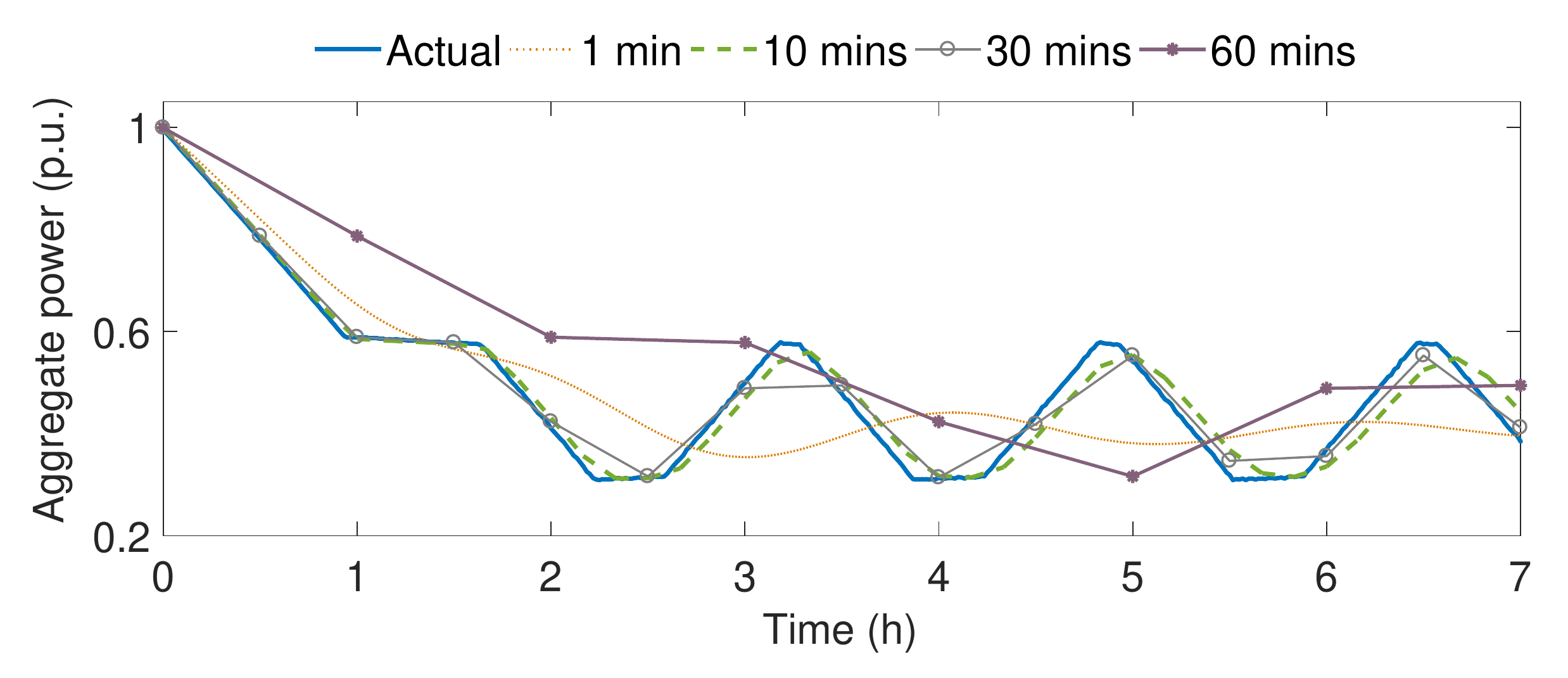}
		\caption{Aggregate demand profiles, for varying market clearing intervals, with $\pi^{\textrm{clr}} = 10$ \$/MWh (Initial temperatures uniform within 19-21$^\textrm{o}$C).}
		\label{fig:VaryTinterval}
	\end{center}	
\end{figure}

\begin{table*}
	\caption{Performance Comparison}	\label{table2}
	\begin{tabularx}{\textwidth}{l*{9}{>{\centering\arraybackslash}X}}
			\hline
											 	&  Case 1   & Case 2   	& Case 3  	&  Case 4   & Case 5   	& Case 6 	& Case 7  	&  Case 8     \\ \hline
			Set up								&    &   	&   	&   &   	&  	&   	&       \\ \hline
			MPC type   	 		  	    		&  MIP   & MIP  & MIP  	&  MIP   &  QP   	& QP 	& QP 	&  QP    	 	\\ 
			Horizon, $N_\textrm{k}$  	  				&  12   & 12    & 18  	&  18   & 18   	& 18 	& 18  	&  18   	\\
			$b^{\textrm{max}}$		     		&  1 & 0.25   	& 1  	&  1    & 0.25 & 0.15 & 0.25 &  0.25  	\\  
			Tuning parameter, $\mu_{\textrm{s}}$		&  0   &  0 	&  0 	&  1000   & 1000   	& 1000 	& 1000  	&  1000    	\\  
			Tuning parameter, $\mu_{\textrm{w}}$	    &  1    & 1   	& 1  	&  1  & 200   & 200 	& 200  	&  200   	\\  \hline
			Results								&    &   	&   	&   &   	&  	&   	&       \\ \hline
			Average system demand, $\mean{D}_k$, in MW 	 			&  4.82   & 7.10  & 4.74  	&  6.08   & 4.68   	& 4.74 	& 4.68  	&  4.68  	\\ 
			Average TCL demand, $\mean{D}^{\textrm{c}}_k$, in MW  			&  1.30  & 3.87   	& 1.28  	&  2.62   & 1.21   	&  1.28	& 1.21 	&  1.21  	\\
			Peak system demand, $\widehat{D}_k$, in MW	  		&  5.20   & 8.00   	& 5.19  	&  8.00   & 5.20  	& 6.78 	& 5.21  	&  5.21 	\\  
			RMSE (normalized), in \%          &  0.82    & 1.19   	& 0.79   	& 0.72	 &  3.22   & 7.88  	&  3.16 	& 3.89 	\\  
		    $\bar{\lambda}^{\textrm{elec}}_k$ (average electricity price), in \$/MWh	&  34.1   & 47   	& 33.7  &  40.4   &  33.4  	& 33.7 	& 33.4  	&  33.4   	\\ 
			$\lambda^{\textrm{elec-}}_k$, $\lambda^{\textrm{elec+}}_k$ (min, max), in \$/MWh	&  30.2, 36.5   & 38.5, 50   &  30, 35.9 	& 26.7, 50.2  & 30.7, 34.8  	& 29.4, 35.8	& 30.8, 34.6 	& 30.7, 34.8  	\\ 		
			Maximum TCLs in a single bin, in \%	&  0.41   & 0.25   	&  0.39 	& 0.16    & 0.25   	& 0.15 	& 0.25  	&  0.25  	\\ 
			Bin spread at $X_{N_\textrm{k}}$			&  5   & 13   	&  5 	& 12    & 12   	& 11 	& 12  	&  12  	\\ 	
			\hline	
	\end{tabularx}
\end{table*}

The effect of changing the market clearing interval $\tau$ is analyzed
next. Consider $\tau = 1$, 10, 30 and 60~minutes. The $A$ matrices,
with $N_{\textrm{B}} = 40$, were obtained for each case under
$\pi^{\textrm{clr}} = 10$~\$/MWh. As shown in
Fig.~\ref{fig:VaryTinterval}, with $\tau = 10$~min, the profile
obtained by \eqref{eq:HybridModel} deviates from the actual. This is
because with a smaller $\tau$, the small changes in temperature (or
bids) can only be accurately captured by using large $N_{\textrm{B}}$
\cite{Bashash2011, Nazir2017a, Mathieu2013}, and using $N_{\textrm{B}}
= 40$ is not sufficient. The profile obtained with $\tau = 10$,
30~minutes match the actual behavior reasonably well. With $\tau =
60$~minutes, the profile again deviates significantly. During the duration
of 60~minutes, many TCLs reach their temperature thresholds and change
state. Hence, the 60~minute bin model could not capture the intra-hour
power consumption dynamics.

\subsection{MPC Performance}

Consider $\tau = 10$~min, and $N_{\textrm{B}} = 20$. Cost of supply,
$C^s_{k}(P^{\textrm{s}}_k) = 10P^{\textrm{s}}_k+ 2.5
(P^{\textrm{s}}_k)^2$. The substation feeder limit is set at 8~MW.

For a storage based system, deciding only based on current or
near-term situations may lead to significant reduction of the feasible
operating region in future periods \cite{Nazir2016,
  Lygeros2015}. Hence, we look ahead several periods in the
MPC. Consider $N_\textrm{k} = 18$. With $\tau = 10$~min, the MPC looks
ahead a 3~hour window (2~hours in cases~1 and~2). Additionally, let
$\sum_{k=1}^{N_\textrm{k}} D^{c}_k \ge 1.21N_\textrm{k}$, where
1.21~MW is an average aggregate TCL demand allowed to avoid the
depletion of the aggregate SOC.

Recall that MPC requires the $A$-matrix. To obtain the coefficients of
$A$, distribute TCLs uniformly over all bins. For TCLs originating
from each sending bin, find the range of bins covered at the end of
$\tau= 10$~minutes. Normalizing these quantities gives the transition
probabilities. Recall that this matrix is not a function of the
clearing price, hence is valid under any clearing price signal.

\begin{figure}[t!]
	\begin{center}
		\includegraphics[scale=0.285]{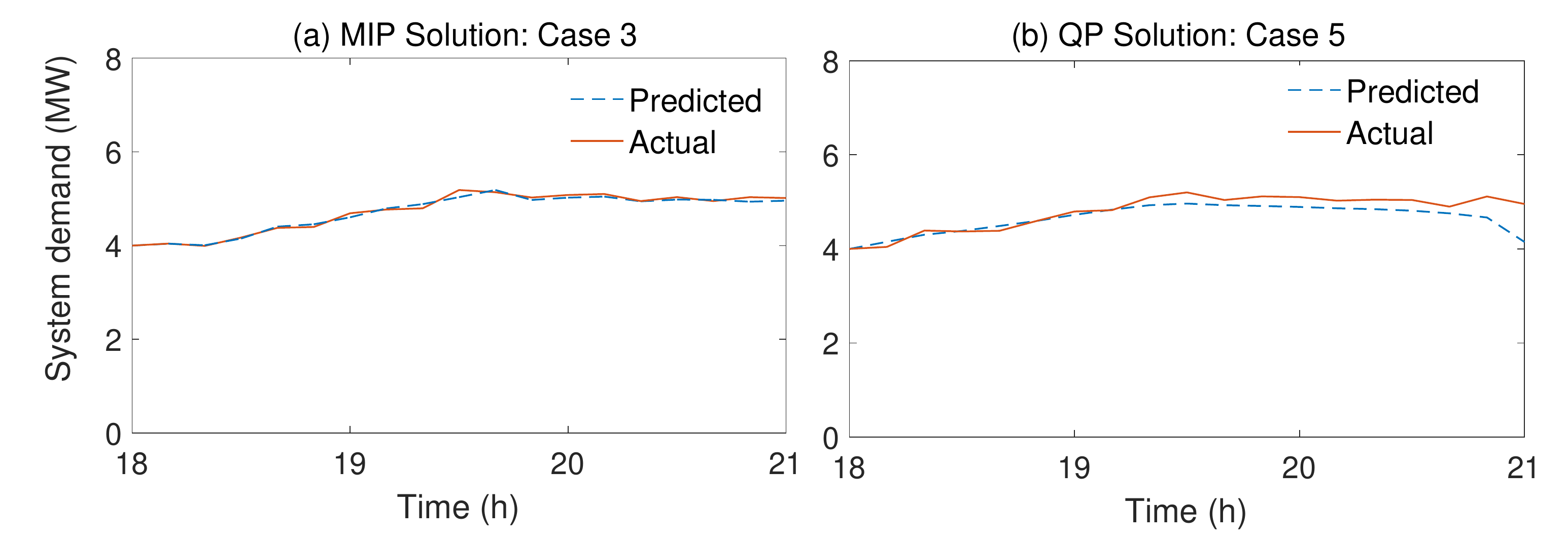}
		\caption{Predicted and actual system demand profiles using (a) MIP in $\quad$ case 3, and (b) QP in case 5.}
		\label{fig:Compare1}
	\end{center}	
\end{figure}

Both the QP and MIP problems were programmed in MATLAB and YALMIP
\cite{Lofberg2004}. QP problems (cases 5-8) were solved using
Quadprog, whereas the MIPs (Cases 1-4) were solved using Gurobi. For
validation, in each case we ran the full dispatch of 1473~TCLs to
test the performance of the MIP/QP solutions. Due to space limitations,
only results for tests starting from hour~18 have been presented. Test
parameters and results are summarized in Table~\ref{table2}. The
average system demand, $\mean{D}_k$, average TCL demand,
$\mean{D}^{\textrm{c}}_k$, and system peak demand $\widehat{D}_k$ are
shown. Also, the average, minimum and maximum price of electricity,
$\bar{\lambda}^{\textrm{elec}}_k$, $\lambda^{\textrm{elec-}}_k$, and
$\lambda^{\textrm{elec+}}_k$, were recorded. The error between the
scheduled and the actual system demand during the dispatch process is
captured by the root mean square error (RMSE) (here, normalized by the
system peak capacity of 8~MW). Initially, temperatures of TCLs were
distributed uniformly within 20 to 21$^\textrm{o}$C. To measure
synchronism, maximum TCLs in a single bin (\%) in $X_k, \forall k$,
and the bin spread (i.e. the number of bins with non-zero quantities
of TCLs) in $X_{N_\textrm{k}}$ were recorded. The following
observations are made,
\begin{itemize}[leftmargin=*]\itemsep1pt 
\item In cases 1 and 3, $b^{\textrm{max}} \le 1$ (i.e. all TCLs can be
  in a single bin) and $\mu_{\textrm{s}} = 0$, hence the risk of
  synchronization was not accounted for. As a result,
  $X_{N_\textrm{k}}$ was narrowly distributed over 5~bins only. Also,
  the maximum fraction of TCLs in a single bin reached approximately
  40\%.
\item Measures to avoid synchronization were taken by setting
  $b^{\textrm{max}} \le 0.25$ in case~2, and by $\mu_{\textrm{s}} =
  1000$ in case~4. Maximum TCLs lying in a single bin decreased
  considerably and wider $X_{N_\textrm{k}}$ were obtained. However,
  $\mean{D}^c_k$, $\widehat{D}_k$ and
  $\bar{\lambda}^{\textrm{elec}}_k$ were significantly higher in
  cases~2 and~4.
\item In case~5, applying the QP solution, the obtained
  $\mean{D}^{\textrm{c}}_k$, $\widehat{D}_k$ and
  $\bar{\lambda}^{\textrm{elec}}_k$ were similar to the values in
  case~3. The demand profiles (and $\lambda^{\textrm{elec}}_k$)
  obtained in cases~3 and~5 are shown in
  Fig.~\ref{fig:Compare1}. However, the RMS error was higher in
  case~5. This is because the QP does not need to enforce strict
  on/off decisions for the entire bin, whereas the MIP does. Under QP,
  contents in a bin can be fractionally chosen to be on. However,
  during dispatch, all TCLs in that bin are cleared due to receiving
  the same price signal. This causes the actual profile to slightly
  deviate from the predicted one.
\item In case~6, when a stricter $b^{\textrm{max}}$ limit was imposed
  than in case~5, $\mean{D}^{\textrm{c}}_k$ and
  $\bar{\lambda}^{\textrm{elec}}_k$ slightly increased. RMSEs and
  $\widehat{D}_k$, however, increased noticeably. Again, this was
  mostly due to the QP solutions favoring fractional ON quantities in
  a bin in order to meet lower $b^{\textrm{max}}$, which eventually
  led to higher error during the dispatch.
\item In case~7, noise (uniform~$[-0.02,0.02]$, in $^\textrm{o}$C/min) was
  introduced in the model, affecting the temperature dynamics of the
  TCLs. The bin model was identified under noisy data. During the
  dispatch process, the RMSE did not increase, rather results were
  comparable to case~5.
\item In case~8, bid slopes were heterogeneous (uniform~$[36, 44]$, in
  \$/MWh). While the cost of the solution
  (i.e. $\bar{\lambda}^{\textrm{elec}}_k$) remained nearly the same,
  the RMSE increased slightly. This can be attributed to the bin
  model's reduced accuracy to deal with parameter heterogeneity. Model
  performance could worsen further when considering heterogeneity in
  other TCL parameters \cite{Nazir2017a}. This could be better
  modeled by using clusters of homogeneous groups \cite{Nazir2017a},
  \cite{Alizadeh2015}.
\end{itemize}
Additionally, it was noted that the average time taken to solve the
MPCs were 95~s for MIPs with $N_\textrm{k} = 18$, 23~s for MIPs with
$N_\textrm{k} = 12$, and less than 2~s for all QP problems.

\subsection{Discussions}

Comparing the MIP and QP solutions in Table~\ref{table2}, in general,
we observe that the average TCL demand, peak demand and average
electricity prices with QP were lower compared to the results obtained
by the MIP\@. However, the QP solutions typically lead to higher error
(i.e. RMSEs values). To reduce the error during the dispatch process,
alternative dispatch schemes could be sought in future
work. Discriminatory prices or incentive signals could be sent to the
TCLs. For example, each TCL could receive a clearing price that is
slightly perturbed by noise. This could potentially reduce the RMSE,
however, detailed investigation should be carried out to analyze the
effectiveness and the fairness of such schemes.

To reduce the likelihood of an oscillatory response and to lower error
during the dispatch, we included $\mu_{\textrm{s}} > 0$ in the
objective or imposed $b^{\textrm{max}}$. Due to the large degree of
freedom when using the QP formulation, the above strategies typically
did not increase the cost of the solution (except in case~6). The more
restrictive MIP form typically led to higher cost solutions,
especially for high $\mu_{\textrm{s}}$ or low $b^{\textrm{max}}$.

Our framework was shown to effectively relieve congestion at a
substation feeder by looking ahead several hours. While for slowly
varying systems, such as EVs and commercial building HVAC systems,
hourly time steps have been used in several recent studies
\cite{Li2014, Bai2017, Hanif2017a, Hanif2017b}, 10~minute price signals
have been used to capture the dynamics of TCLs in our case. Future
work could apply our proposed framework in a rolling horizon setting
and for other power system applications. One natural extension could
be to deal with network congestion. Recent work investigates
distribution locational marginal prices (LMPs) to alleviate network
congestion while considering dynamics of EVs \cite{Li2014, Bai2017}
and HVACs \cite{Hanif2017a, Hanif2017b}. Compared to these approaches,
our method has the advantage that it lets users choose their
individual bid functions and the aggregator makes decisions utilizing
the bin-based aggregate model. One could consider aggregating loads at
different nodes of the network, which will then allow computing the
nodal LMPs and the optimal incentive signals for each aggregation.

\section{Conclusions} \label{sec:concl}

Recent studies have shown that transactive coordination of DERs, such
as TCLs, batteries, and EVs, can lead to undesirable power
oscillations. In this paper, we analyzed the causes of such
oscillations and identified different factors that can affect the
aggregate DER dynamics. The user defined bid slopes, preference for
setting locking conditions, price signals sent to DERs for
coordinating their responses, and imposing feeder limits, all can
affect the natural charging and discharging cycles of DERs, hence can
lead to load synchronization and undesired power oscillations. To
address these issues, we developed a bin-based DER aggregate model
under transactive coordination. The trade-offs when using varying
model orders and market intervals were analyzed. With reformulation of
the transition equations, we showed how the model can be incorporated
in an MPC framework. The MPC can then be solved to find optimal price
signals that should be sent to the DERs for governing their aggregate
responses in a desired manner. An accurate MIP formulation, and a
relaxed QP formulation have been developed and tested. Simulation
results compared the performance under several different scenarios by
varying the initial conditions, penalty for synchronization, noise and
heterogeneity. Future work could involve extending our approach to
study hierarchical coordination among system operators, DSOs and
aggregators. Schemes that can consider non-uniform incentive signals
could also be investigated.

\section{Appendices}

\subsection{TCL Model}\label{sec:TCLmodel}

The internal and ambient temperatures ($^\circ$C) corresponding to
each TCL load $j$ are denoted by $\theta_j$ and $\theta^{\textrm{a}}$,
respectively. Each load can be modeled as a thermal capacitance,
$C_j$~(kWh/$^\circ$C), in series with a thermal resistance,
$R_j$~($^\circ$C/kW). The binary variable $m_j$ denotes the on/off
status of the load, and $P_j$~(kW) is the energy transfer rate when a
cooling (or heating) TCL is switched ON\@. The dynamics of a TCL's
temperature can be modeled using a first-order difference equation
\cite{Malhame1985}, \cite{Mortensen1988},
\begin{equation} \label{eq:TCL_DE}
\theta_{j, k+1} = \tilde{a}_{j}\theta_{j,k}+(1-\tilde{a}_{j}) (\theta^{\textrm{a}}-m_{j,k}
\theta^{\textrm{g}}_{j}),
\end{equation}
where the time between $k$ and $k+1$ is $h$ (in our simulation, $h =
10$~s), $\tilde{a}_j = e^{\nicefrac{-h}{(C_j R_j)}}$ is the
parameter governing the thermal characteristics of the thermal mass,
and $\theta^{\textrm{g}}_{j}= P_j R_j$ is the temperature gain when a
cooling TCL is ON\@. The variable $m_{j,k}$ switches off when
$\theta_{j,k}<\theta^{\textrm{min}}_{j}$ and on when $\theta_{j,k}>
\theta^{\textrm{max}}_{j}$. Comparing with the battery model in
\eqref{eq:GBM1}, $e_j = 1$ when $\theta_{j,k}$ reaches
$\theta^{\textrm{min}}_{j}$ and $e_j = 0$ when $\theta_{j,k}$ reaches
$\theta^{\textrm{max}}_{j}$.  With coefficient of performance (COP)
$\eta_j$, the electrical power consumed is given by,
$P^{\textrm{elec}}_{j,k} = {\nicefrac{m_{j,k} P_j}{ \eta_j}}$.

Following the calculation procedure in \cite{Callaway2009}, for a
176~m$^2$ house with 3~ton (approximately 10.55~kW) AC
\cite{Perez2014} and $\eta_j = 3.5$, the parameters would be
$P_j^{\textrm{elec}} = 3$~kW when on, $R_j = 2.84^\circ$C/kW, and $C_j
= 7.04$~kWh/$^\circ$C.



%
%

\ifCLASSOPTIONcaptionsoff
  \newpage
\fi



%

\bibliographystyle{IEEEtran}
\bibliography{PEStransactive1}

%
%
%
%
%

%

\begin{IEEEbiography}[{\includegraphics[width=1in,height=1.25in,clip,keepaspectratio]{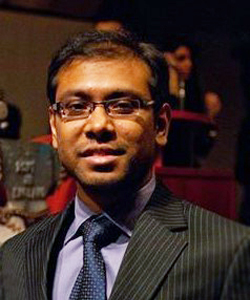}}]{Md Salman Nazir}
(S’10-M’13) received his B.Eng. (Honours) and M.Eng. (Thesis) degrees, both in Electrical Engineering, from McGill University, Montreal, Canada in 2011 and 2015, respectively. He is currently pursuing his Ph.D. degree in Electrical and Computer Engineering at the University of Michigan, Ann Arbor, MI, USA. During 2012-2014, he was an engineer at the Natural Resources Canada's CanmetENERGY laboratory. His research interests include operations, control and economics of power systems in the presence of significant renewable and distributed energy resources.
\end{IEEEbiography}
%
\vspace{-5mm}
\begin{IEEEbiography}[{\includegraphics[width=1in,height=1.25in,clip,keepaspectratio]{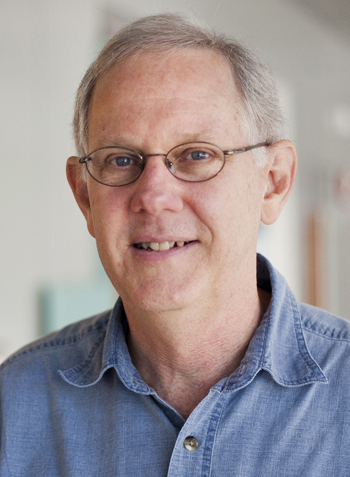}}]{Ian A. Hiskens}
(F’06) is currently the Vennema Professor of Engineering with the Department of
Electrical Engineering and Computer Science, University of Michigan,
Ann Arbor, MI, USA. He has held prior appointments in the Queensland
electricity supply industry, and various universities in Australia
and the United States. His research interests lie at the
intersection of power system analysis and systems theory, with
recent activity focused largely on integration of renewable
generation and controllable loads. Dr. Hiskens is actively involved
in various IEEE societies and was VP-Finance of the IEEE Systems
Council. He is a Fellow of IEEE, a Fellow of Engineers Australia and a Chartered
Professional Engineer in Australia.
\end{IEEEbiography}

%




\end{document}